\documentclass[aps,prd,preprintnumbers,groupedaddress,nofootinbib,amssymb,eqsecnum,longbibliography,10pt]{revtex4-2}

\usepackage{graphicx}
\usepackage{color} 
\usepackage{dcolumn}
\usepackage{amsmath,amsthm,amssymb}
\usepackage{amsfonts}
\usepackage{bm}
\usepackage{mathrsfs}
\usepackage{revsymb}
\usepackage{comment}
\usepackage{braket}
\usepackage{simplewick} 

\usepackage[bookmarks=true,bookmarksnumbered=true,colorlinks=false]{hyperref}
\newif\ifoneauthor
\oneauthortrue

\newcommand{\cc}{{0}}

\begin{document}

\title{Quantum decoherence of gravitational waves}

\author{Hiroki Takeda}
\email[]{takeda@tap.scphys.kyoto-u.ac.jp}
\affiliation{The Hakubi Center for Advanced Research, Kyoto University, Kyoto 606-8501, Japan}
\affiliation{Department of Physics, Kyoto University, Kyoto 606-8502, Japan}
\author{Takahiro Tanaka}
\affiliation{Department of Physics, Kyoto University, Kyoto 606-8502, Japan}
\affiliation{Center for Gravitational Physics and Quantum Information, Yukawa Institute for Theoretical Physics, Kyoto University, Kyoto 606-8502, Japan}

\date{\today}

\begin{abstract}
\noindent
The quantum nature of gravity remains an open question in fundamental physics, lacking experimental verification. 
Gravitational waves (GWs) provide a potential avenue for detecting gravitons, the hypothetical quantum carriers of gravity. 
However, by analogy with quantum optics, distinguishing gravitons from classical GWs requires the preservation of quantum coherence, which may be lost due to interactions with the cosmic environment causing decoherence.
We investigate whether GWs retain their quantum state by deriving the reduced density matrix and evaluating decoherence, using an environmental model where a scalar field is conformally coupled to gravity.
Our results show that quantum decoherence of GWs is stronger at lower frequencies and higher reheating temperatures.
We identify a model-independent amplitude threshold below which decoherence is negligible, providing a fundamental limit for directly probing the quantum nature of gravity.
In the standard cosmological scenario, the low energy density of the universe at the end of inflation leads to complete decoherence at the classical amplitude level of inflationary GWs.
However, for higher energy densities, decoherence is negligible within a frequency window in the range $100\ {\rm Hz} \text{--} 10^8\ {\rm Hz}$, which depends on the reheating temperature.
In a kinetic-dominated scenario, the dependence on reheating temperature weakens, allowing GWs to maintain quantum coherence above $10^7\ {\rm Hz}$.
\end{abstract}

\maketitle


\section{Introduction}
Gravity, while the weakest of the fundamental forces, continues to pose profound mysteries in fundamental physics. 
Despite its subtlety, gravity has revealed to humanity that celestial and terrestrial phenomena are governed by the same laws, driving countless advances from the development and testing of general relativity to the observation of gravitational waves (GWs)~\cite{Will:2014kxa, KAGRA:2021vkt, LIGOScientific:2021sio}. 
The unification of gravity and quantum theory remains one of the grand ambitions in physics, as it is expected to shed light on fundamental questions such as the evolution of spacetime and matter at the Planck scale, while perturbatively quantized general relativity can be regarded as a low-energy effective quantum field theory without any inconsistencies~\cite{Donoghue:1994dn, Donoghue:1995cz, Burgess:2003jk, Donoghue:2022eay}. 
However, its inherent weakness makes uncovering the nature of gravity challenging, and there is still no experimental evidence that gravity is quantized. 
As a counterargument, theories treating gravity as a semi-classical or classical phenomenon have also been considered~\cite{Moller:1959bhz, Rosenfeld:1963hjy, Page:1981aj, Blanchard:1994se, Diosi:1995qs, Oppenheim:2018igd}. 

There are two main approaches to experimentally probing the quantumness of gravity. 
The first involves exploring scenarios in which spacetime itself can exist in a superposition of macroscopically distinct states~\cite{DeWitt:1957obj, Bose:2017nin, Marletto:2017kzi, Carney:2018ofe, Aspelmeyer:2022fgc}. 
Quantum gravitational interactions can mediate entanglement between macroscopic quantum systems, whereas classical gravitational fields cannot induce such entanglement between quantum systems. 
Therefore, if coherent superpositions of macroscopic objects can be created, resulting in an observable entangled state through their gravitational interaction, it would provide experimental evidence for the quantum nature of gravity. 
Several experimental setups have been considered that use massive particles~\cite{Bose:2017nin, Marletto:2017kzi, Pedernales:2021dja}, optomechanical systems~\cite{Krisnanda:2017sey, Qvarfort:2018uag, Miao:2019pxw, Datta:2021ywm}, and atoms~\cite{Haine:2018bwu, Carney:2021yfw,Howl:2023xtf}.
However, the gravitational interaction responsible for generating entanglement is extremely weak, and confirming the quantum behavior of macroscopic objects is inherently challenging~\cite{Wood:2021icq, Aspelmeyer:2022fgc, Miki:2024qcz}. 
Realizing coherent quantum states of sufficiently massive objects, generating detectable entanglement through weak gravitational interactions, and measuring this entanglement while eliminating classical noise and avoiding decoherence due to disturbances are significant practical challenges. 
Moreover, even if gravitationally induced entanglement were observed, it would not directly confirm the quantization of gravity, as even Newtonian gravity, which lacks independent gravitational degrees of freedom, can introduce quantum entanglement between matter systems.

Another approach is a more direct method of probing gravitons, the hypothetical quantum carriers of gravity. 
The possibility of directly detecting gravitons has been explored since the early works of~\cite{Dyson:2013hbl, Boughn:2006st, Rothman:2006fp}.
Since the efficiency of gravitational detectors is extremely low compared to modern photo-counting devices,
recent proposals focus on developing more effective methods to establish the existence of gravitons by utilizing quantum noise induced by gravitons~\cite{Calzetta:1993qe, Anastopoulos:2013zya, Oniga:2015lro, Kanno:2020usf, Parikh:2020nrd} or quantum sensing techniques~\cite{Tobar:2023ksi}.
However, as suggested by analogies with quantum optics, clearly distinguishing between classical and quantum behavior requires GWs to exist in non-trivial quantum states, such as squeezed states~\cite{Guerreiro:2021qgk, Carney:2023nzz}.
Consequently, how detectors respond to quantum states of GWs has been investigated~\cite{Pang:2018eec, Kanno:2018cuk, Carney:2023nzz}.
Unlike photons, generating non-trivial quantum states of gravitons is considered extremely challenging. 
Therefore, the only plausible context where such quantum states might naturally emerge would be during inflation. 
In fact, some models predict that GWs may exist in quantum states that are distinguishable from classical signals~\cite{Grishchuk:1989ss, Grishchuk:1990bj, Albrecht:1992kf, Polarski:1995jg}. 
However, GWs originating from inflation may lose their quantum coherence due to interactions with environmental matter before reaching the detectors.
In this paper, we evaluate the quantum decoherence of GWs to address the question: {\it Can gravitons retain their quantum nature?} This is crucial for definitively establishing the existence of gravitons as quantum entities, completely distinguishing them from classical signals, even if humanity were to devise a highly efficient detection method. Moreover, understanding the quantum nature of GWs could provide valuable insights into the generation and evolution of primordial tensor perturbations, which are key to probing the mechanism of cosmic inflation.

In Sec.~\ref{sec:formulation}, we provide the reduced density matrix using the influence functional for a model in which GWs, serving as the system, are conformally coupled to a scalar matter field as the environment on the FLRW background.
In Sec.~\ref{sec:master_equation}, we introduce the noise and dissipation kernels through a perturbative expansion.
In Sec.~\ref{sec:decoherence_functional}, we derive a decoherence functional characterizing the quantum decoherence of GWs.
In Sec.~\ref{sec:decoherence_function}, we discuss the quantum decoherence of the primordial GWs by evaluating the decoherence function for cosmic expansion histories.
Finally, we conclude the paper in Sec.~\ref{sec:conclusion}.
Throughout this paper, we adopt natural units where $c=\hbar=k_B=1$ unless explicit dependence is specified, and define the reduced Planck mass as $M_{\rm pl}:= (8\pi G)^{-1/2}$ where $G$ denotes the gravitational constant. 
We also adopt the metric signature $(+, -, -, -)$.

\section{Formulation}
\label{sec:formulation}
We consider an action of the form $S[g_{\mu\nu}, \phi]=\int d^4x \,\mathcal{L}(g_{\mu\nu}, \phi, \nabla_{\alpha}\phi)$ 
where the Lagrangian density $\mathcal{L}$ includes the Einstein-Hilbert term and the scalar field contribution $\mathcal{L}=\mathcal{L}_{\rm EH}+\mathcal{L}_{\rm \phi}$. The Einstein-Hilbert action is given by
\begin{align}
    S_{\rm EH}=\int d^4x\mathcal{L}_{\rm EH}=\int d^4x \sqrt{-g}\frac{1}{2\kappa}R\,,\label{eq:S_EH}
\end{align}
where $\kappa:=8\pi G$ and $R$ is the Ricci scalar.
The scalar field action is given by
\begin{align}
    S_{\phi} = \int d^4x \mathcal{L}_{\phi} = \int d^4x \sqrt{-g} \left( -\frac{1}{2}g^{\mu\nu}\nabla_{\mu}\phi\nabla_{\nu}\phi -\frac{1}{2}m^2\phi^2 -\frac{1}{2}\xi R \phi^2\right)\,,\label{eq:S_phi}
\end{align}
where $m$ is the mass of a conformally coupled scalar field $\phi$ with $\xi = 1/6$.
The scalar field $\phi$ serves as a proxy for environmental matter degrees of freedom that may cause decoherence of the GWs
\footnote{Although $\phi$ is a conformally coupled scalar for simplicity, it can be viewed as a stand-in for more realistic matter fields such as neutrinos, photons, or any matter fields weakly interacting with gravity.} 
.

Let us consider a linear metric perturbation around the Friedmann-Lemaître-Robertson-Walker (FLRW) metric 
\begin{align}
    ds^2 = a^2(\eta)\Big[ d\eta^2 - (\delta_{ij} + h_{ij}) dx^i dx^j\Big]\,,
\end{align}
where $a$ is the scale factor, $\eta$ is the conformal time defined by $dt = a(t) d\eta$, and we use $x := (x^0, {\bm x})=(\eta, {\bm x})$ for the 4-dimensional coordinates.
Here, $h_{ij}$ is the transverse-traceless tensor satisfying $\partial^{j}h_{ij}=0$ and $h_{i}{}^{i}=0$.
Since we consider a linear perturbation of the scalar field around a zero background, we redefine the linear perturbation as $\phi$.
By perturbing Eq.~\eqref{eq:S_EH} and Eq.~\eqref{eq:S_phi}, we find the action up to the third order perturbation,
\begin{align}
S[h_{ij}, \chi] = S_{0}[h_{ij}] + S_{0}[\chi] + S_{\rm int}[h_{ij}, \chi]\,,
\end{align}
where
\begin{align}
    S_{0}[h_{ij}] &= \frac{1}{2\kappa} \int d^4x \frac{1}{4}a^2 \left(\partial_{\eta}h_{ij} \partial_{\eta}h^{ij} - \partial_{k}h_{ij}\partial^{k}h^{ij}\right)\,,
\end{align}
is the free part of the GWs,
\begin{align}
S_{0}[\chi]&= \int d^4x \left[\frac{1}{2}(\partial_{\eta} \chi)^2 - \frac{1}{2}(\partial_{i}\chi)^2 -\frac{1}{2}m_{\rm eff}^2\chi^2\right]\,,
\end{align}
is the free part of the scalar field $\chi:=a\phi$ where $m_{\rm eff}^2 := a^2(\eta) m^2$ is the effective mass of the scalar field $\chi$, and
\begin{align}
S_{\rm int}[h_{ij}, \chi] = \int d^4x \left[\frac{1}{2}h^{ij}\partial_{i}\chi \partial_{j}\chi\right]\,,
\end{align}
is the interaction action.

We define the Fourier transform as
\begin{align}
    \phi(x) = \frac{1}{(2\pi)^4} \int d^4k\ \tilde{\phi}(k)e^{ik_{\mu}x^{\mu}}\,,\quad
    h_{ij}(x) = \frac{1}{(2\pi)^4} \int d^4k\ \tilde{h}_{ij}(k)e^{ik_{\mu}x^{\mu}}\,,
\end{align}
where $k = (k^0, {\bm k})$ is the four momentum conjugate to $x$.
Since $\phi$ and $h_{ij}$ are real valued, we have conditions on the Fourier modes $\tilde{\phi}^{*}(k) =\tilde{\phi}(-k)$ and $\tilde{h}_{ij}^{*}(k) =\tilde{h}_{ij}(-k)$.
We also decompose $\phi$ and $h_{ij}$ into the spatial Fourier modes
\begin{align}
    \phi(\eta, {\bm x}) = \frac{1}{(2\pi)^3} \int d^3k\ \phi_{\bm k}(\eta)e^{i{\bm k}\cdot {\bm x}}\,,\quad
    h_{ij}(\eta, {\bm x}) = \frac{1}{(2\pi)^3} \int d^3k\ {h}_{ij, {\bm k}}(\eta)e^{i{\bm k}\cdot{\bm x}}\,.
\end{align}
For the reality of $\phi$ and $h_{ij}$, we have $\phi_{\bm k}^{*}(\eta) =\phi_{-{\bm k}}(\eta)$ and ${h}_{ij, {\bm k}}^{*}(\eta) ={h}_{ij, -{\bm k}}(\eta)$.
By introducing the polarization basis $e_{ij}^A(\hat{\bm k})$, we further decompose the spatial Fourier components into the mode functions,
\begin{align}
    h_{ij, {\bm k}}(\eta) = \sum_{A=+, \times} e_{ij}^A(\hat{\bm k}) h_{\bm k}^{A}(\eta)\,,
\end{align}
where $A$ is the polarization label running over plus and cross polarization modes.
Since we have $e_{ij}(-\hat{\bm k}) = e_{ij}^{*}(\hat{\bm k}) = e_{ij}(\hat{\bm k})$  if we set the conventional orthonormal GW basis, the condition for the reality of $h_{ij}$ reduces to ${h}^{A*}_{\bm k}(\eta)={h}^{A}_{-\bm k}(\eta)$.
In a similar manner, we also write the Fourier modes in the polarization basis as
\begin{align}
    \tilde{h}_{ij}(k) = \sum_{A=+, \times} e_{ij}^A({\hat{\bm k}})\tilde{h}^{A}(k)\,,
\end{align}
where $\tilde{h}^{A}(k)$ is related to the mode function $h_{\bm k}^{A}(\eta)$ by Fourier transform
\begin{align}
     h_{\bm k}^{A}(\eta) = \frac{1}{2\pi}\int dk_0\ \tilde{h}^{A}(k)e^{-ik_0 \eta}\,.
\end{align}

We quantize the scalar field $\chi$ by promoting it to an operator,
\begin{align}
    \hat{\chi}(x) = \frac{1}{(2\pi)^3}\int d^3k \left[\hat{a}_{\bm k}\chi_{\bm k}(\eta)e^{i{\bm k}\cdot{\bm x}} +  \hat{a}_{\bm k}^{\dagger}\chi^{*}_{\bm k}(\eta)e^{-i{\bm k}\cdot{\bm x}}\right]\,,
\end{align}
where $\hat{a}_{\bm k}$ and $\hat{a}^{\dagger}_{\bm k}$ are the annihilation and creation operators for the mode with momentum ${\bm k}$ satisfying the commutation relation
\begin{align}
    [\hat{a}_{\bm k}, \hat{a}^{\dagger}_{\bm k'}]=(2\pi)^3 \delta^{(3)}({\bm k}-{\bm k'})\,,
\end{align}
and $\chi_{\bm k}(\eta)$ is the mode function determined by solving the equation of motion.
In Heisenberg representation, we define the eigenstate $\ket{\chi, \eta}_{\rm H}$ of the field operator $\hat{\chi}(x)$ as
\begin{align}
    \hat{\chi}(x)\ket{\chi, \eta}_{\rm H} = \chi(x)\ket{\chi, \eta}_{\rm H}\,.
\end{align}
We also quantize $h_{ij}(x)$ by promoting it to an operator $\hat{h}_{ij}(x)$,
\begin{align}
    \hat{h}_{ij}(x) = \frac{1}{(2\pi)^3} \int d^3k\  \sum_{A=+, \times} e_{ij}^A(\hat{\bm k}) \left[\hat{b}_{\bm k}^{A} h_{\bm k}^{A}(\eta)e^{i{\bm k}\cdot{\bm x}} + \hat{b}^{A \dagger}_{\bm k} h_{\bm k}^{A*}(\eta)e^{-i{\bm k}\cdot{\bm x}}\right]\,,
\end{align}
where $\hat{b}^{A}_{\bm k}$ and $\hat{b}^{A \dagger}_{\bm k}$ are the annihilation and creation operators for the mode with A and ${\bm k}$, which satisfies the commutation relation
\begin{align}
    [\hat{b}^{A}_{\bm k}, \hat{b}^{B}_{\bm k'}] = (2\pi)^3\delta^{(3)}({\bm k}- {\bm k'})\delta_{AB}\,.
\end{align}
In Heisenberg representation, the eigenstate $\ket{h_{ij}}$ of the field operator $\hat{h}_{ij}(x)$ is defined by
\begin{align}
    \hat{h}_{ij}(x)\ket{h_{ij},\eta}_{\rm H} = h_{ij}(x)\ket{h_{ij}, \eta}_{\rm H}\,.
\end{align}
For simplicity, when no confusion arises, we will denote the eigenstate $\ket{h_{ij}, \eta}_{\rm H}$ of the gravitational field operator $\hat{h}_{ij}(x)$ as $\ket{h, \eta}_{\rm H}$. 

The components of the total density matrix is defined by
\begin{align}
    \rho[h^{+},\chi^{+}, h^{-}, \chi^{-}, \eta_f]:={}_{\rm H}\braket{h^{+}, \chi^{+}, \eta_f |\hat{\rho}^{H} | h^{-}, \chi^{-}, \eta_f}_{\rm H}\,,
\end{align}
where $\hat{\rho}^{H}$ is the total density matrix operator and the subscript $+, -$ denotes the closed time path branches.
By tracing out the environment $\chi$, we obtain the reduced density matrix
\begin{align}
    \rho_{r}[h^{+}, h^{-}, \eta_f] := \int d\chi^{+} \rho[h^{+}, \chi^{+}, h^{-}, \chi^{+}, \eta_f]\,.
\end{align}

If we assume that the total density matrix is initially uncorrelated
\begin{align}
    \hat{\rho}^{H} = \hat{\rho}_{h}^{H} \otimes \hat{\rho}_{\chi}^{H} \,,
\end{align}
the reduced density matrix of the system field $h$ evolves in time as~\cite{Calzetta:2008iqa}
\begin{align}
    \rho_{r}[h^{+}_{f}, h^{-}_{f}, \eta_f] = \int dh^{+}_{i} \int dh^{-}_{i} \mathcal{J}_{r}[h^{+}_{f}, h^{-}_{f}, \eta_f ; h^{+}_{i}, h^{-}_{i}, \eta_i] \rho_{h}[h^{+}_{i}, h^{-}_{i}, \eta_i]\,.
\end{align}
where the propagator $\mathcal{J}_{r}[h^{+}_{f}, h^{-}_{f}, \eta_f ; h^{+}_{i}, h^{-}_{i}, \eta_i]$ is given by a closed time path integral
\begin{align}
    \mathcal{J}_{r}[h^{+}_{f}, h^{-}_{f}, \eta_f ; h^{+}_{i}, h^{-}_{i}, \eta_i] = \int_{h^{+}_{i}(\bm{x})}^{h^{+}_{f}(\bm{x})} D h^{+} \int_{h^{-}_{i}(\bm{x})}^{h^{-}_{f}(\bm{x})} Dh^{-} \exp{\left[\frac{i}{\hbar}S_{\rm eff}[h^{+}, h^{-}]\right]}\,.
\end{align}
Here, the full influence functional effective action is given by
\begin{align}
    S_{\rm eff}[h^{+}, h^{-}] =  S_{0}[h^{+}] - S_{0}[h^{-}] + S_{\rm IF}[h^{+}, h^{-}]\,,
\end{align}
where the influence action $S_{\rm IF}$ is given by the Feynman-Vernon influence functional $\mathcal{F}[h^{+}, h^{-}]$ as
\begin{align}
    \mathcal{F}[h^{+}, h^{-}] &= \exp{\left[\frac{i}{\hbar}S_{\rm IF}[h^{+}, h^{-}]\right]}\\
    &= \int d\chi^{+}_{f}({\bm x}) \int d\chi^{+}_{i}({\bm x}) \int d\chi^{-}_{i}({\bm x})  \rho_{\chi}[\chi^{+}_{i}, \chi^{-}_{i}, \eta_{i}] \int_{\chi^{+}(\eta_i ,\bm{x})=\chi^{+}_{i}(\bm{x})}^{\chi^{+}(\eta_f ,\bm{x})=\chi^{+}_{f}(\bm{x})} D \chi^{+} \int_{\chi^{-}(\eta_i ,\bm{x})=\chi^{-}_{i}(\bm{x})}^{\chi^{-}(\eta_f ,\bm{x})=\chi^{+}_{f}(\bm{x})} D\chi^{-}\nonumber \\
    &\times \exp{\left[\frac{i}{\hbar}\left\{ S_{0}[\chi^{+}] - S_{0}[\chi^{-}] +S_{\rm int}[h^{+}, \chi^{+}] - S_{\rm int}[h^{-}, \chi^{-}] \right\}\right]}\,.
\end{align}

We expand the influence action up to second order in the coupling constant $1/\kappa$ and first order in $\hbar$,
we find
\begin{align}
    S_{\rm IF}[h^{+}, h^{-}] &= \braket{S_{\rm int}[h^{+}, \chi^{+}]} - \braket{S_{\rm int}[h^{-}, \chi^{-}]}\nonumber \\
    &+\frac{i}{2\hbar}\left[ \braket{S_{\rm int}[h^{+}, \chi^{+}]^2} - \braket{S_{\rm int}[h^{+}, \chi^{+}]}^2\right] \nonumber \\
    &+\frac{i}{\hbar}\left[ \braket{S_{\rm int}[h^{+}, \chi^{+}]}\braket{S_{\rm int}[h^{-}, \chi^{-}]} - \braket{S_{\rm int}[h^{+}, \chi^{+}]S_{\rm int}[h^{-}, \chi^{-}]}\right]\nonumber \\
    &+\frac{i}{2\hbar}\left[ \braket{S_{\rm int}[h^{-}, \chi^{-}]^2} - \braket{S_{\rm int}[h^{-}, \chi^{-}]}^2\right]\,,
\end{align}
Here, the quantum average of a physical quantity $Q[h^{+}, \chi^{+}, h^{-}, \chi^{-}]$ over the unperturbed action $S_{0}[\chi]$ is defined by
\begin{align}
    \braket{Q[h^{+}, \chi^{+}, h^{-}, \chi^{-}]}
    &= \int d\chi^{+}_{f}({\bm x}) \int d\chi^{+}_{i}({\bm x}) \int d\chi^{-}_{i}({\bm x})  \rho_{\chi}[\chi^{+}_{i}, \chi^{-}_{i}, \eta_{i}] \int_{\chi^{+}(\eta_i ,\bm{x})=\chi^{+}_{i}(\bm{x})}^{\chi^{+}(\eta_f ,\bm{x})=\chi^{+}_{f}(\bm{x})} D \chi^{+} \int_{\chi^{-}(\eta_{i} ,\bm{x})=\chi^{-}_{i}(\bm{x})}^{\chi^{-}(\eta_f ,\bm{x})=\chi^{+}_{f}(\bm{x})} D\chi^{-} \nonumber \\
    &\times \exp{\left[\frac{i}{\hbar}\left\{ S_{0}[\chi^{+}] - S_{0}[\chi^{-}]\right\}\right]} Q[h^{+}, \chi^{+}, h^{-}, \chi^{-}]\,.
\end{align}

\section{Master equation}
\label{sec:master_equation}
We change the field variables to the Keldysh variables defined by
\begin{align}
    h_c &:= \frac{1}{2}\left(h^{+} + h^{-}\right)\,,\quad h_{\Delta} := h^{+} - h^{-}\,,\\
    \chi_c &:= \frac{1}{2}\left(\chi^{+} + \chi^{-}\right)\,,\quad \chi_{\Delta} := \chi^{+} - \chi^{-}\,,
\end{align}
under which the interaction action transform as
\begin{align}
    {S}_{\rm int}[{h}^{c}, {h}^{\Delta}, {\chi}^{c}, {\chi}^{\Delta}] &= -\frac{1}{2\kappa}\int d^{4}k_1 \int d^{4}k_2 \int d^{4}k_3 \sum_{A}\left[\left\{\tilde{h}_{\Delta}^{A}(k_1)\tilde{\chi}_c(k_2)\tilde{\chi}_c(k_3) + \tilde{h}_c^{A}(k_1)\tilde{\chi}_{\Delta}(k_2)\tilde{\chi}_c(k_3)\right.\right.\nonumber \\
    &+ \left.\left.\tilde{h}_c^{A}(k_1)\tilde{\chi}_c(k_2)\tilde{\chi}_{\Delta}(k_3) +\frac{1}{4}\tilde{h}_{\Delta}^{A}(k_1)\tilde{\chi}_{\Delta}(k_2)\tilde{\chi}_{\Delta}(k_3) \right\} e_{ij}^{A}(\hat{\bm k}_{1})k_{2}^{i}k_{3}^{j}\delta(k_{1} + k_{2} + k_{3})\right]\,.
\end{align}

The influence action up to second order in the coupling constant $1/\kappa$ can be rewritten in terms of the Keldysh variables,
\begin{align}
    S_{\rm IF}[h^{+}, h^{-}] &= \braket{{S}_{\rm int}[h^{c},h^{\Delta}, \chi^{c}, \chi^{\Delta}]}_{0} +\frac{i}{2\hbar}\left[ \braket{{S}_{\rm int}[h^{c},h^{\Delta} \chi^{c}, \chi^{\Delta}]^2}_{0} - \braket{{S}_{\rm int}[h^{c}, h^{\Delta}, \chi^{c}, \chi^{\Delta}]}_{0}^2\right]\,.
\end{align}
We decompose the higher point correlations according to Wick's theorem.
The contribution at the tree level disappears because of $\int d^3k\, k^i k^j F(k)\propto \delta^{ij}$ for an isotropic function $F(k)$. 
Thus, the leading contribution at one-loop order is expressed by
\begin{align}
    &\braket{{S}_{\rm int}[h^{c},h^{\Delta}, \chi^{c}, \chi^{\Delta}]^2}-\braket{{S}_{\rm int}[h^{c},h^{\Delta}, \chi^{c}, \chi^{\Delta}]}^2 \nonumber \\
    &=\int d^{4}p \sum_{A}\left[ \tilde{h}^{A}_{\Delta}(p)\tilde{h}^{A}_{\Delta}(-p)\mathcal{N}(p) + i\int d\omega \tilde{h}^{A}_c(p^{0}, {\bm p})\tilde{h}^{A}_{\Delta}(\omega, -{\bm p})\tilde{\mathcal{D}}(p, \omega)\right]\,,
\end{align}
where the noise kernel is given by
\begin{align}
    \mathcal{N}(p) = 2 \int d^{4}k \left[ \bm{k}^2 -\frac{1}{\bm{p}^2}(\bm{p}\cdot\bm{k}) ^2\right]^2 {\rm Re}[G_{F}(k)G_{F}(-p-k)]\,,
    \label{eq:noise_kernel}
\end{align}
and the dissipation kernel takes the form
\begin{align}
    \tilde{\mathcal{D}}(p, \omega) = 2 \int d^{4}k \int d\eta d\eta' d\omega' \left[ \bm{k}^2 -\frac{1}{\bm{p}^2}(\bm{p}\cdot\bm{k}) ^2\right]^2 \Theta(\eta, \eta', p^{0}, \omega, k^{0}, \omega'){\rm Im}[G_{F}(k^{0}, {\bm k})G_{F}(\omega', -{\bm p}-{\bm k})]\,,
\end{align}
with the function $\Theta(\eta, \eta', p^{0}, \omega, k^{0}, \omega')$ defined by
\begin{align}
    \Theta(\eta, \eta', p^{0}, \omega, k^{0}, \omega') := \theta(\eta-\eta')e^{-i(p^{0}+k^{0}+\omega')\eta}e^{-i(\omega-k^{0}-\omega')\eta'}\,.
\end{align}
where $\theta(\eta-\eta')$ is the step function.
Here, the Feynman propagator $G_{F}(k)=G^{0}_{F}(k)+G^{\beta}_{F}(k)$ is separated into two parts, each obtained by substituting the effective mass $m_{\rm eff}$ evaluated at a reference time into the propagators in a Minkowski background:
the free part 
\begin{align}
    G_{F}^0(k) = -i\hbar \frac{1}{-(k^{0})^2+\bm{k}^2+m^{2}_{\rm eff}-i\epsilon}\,,
    \label{eq:free_propagator}
\end{align}
and the thermal part
\begin{align}
    G_{F}^{\beta}(k) = 2\pi n_{B}(k^0) \delta(-(k^{0})^2+\bm{k}^2+m^{2}_{\rm eff})\,,
    \label{eq:thermal_propagator}
\end{align}
where the scalar matter field is bosonic and follows the Bose-Einstein distribution
\begin{align}
n_{B}(k^0) = \frac{1}{e^{\beta(k^0-\mu)}-1}\,.
\end{align}
Here, $\beta =1/T$ is the inverse temperature, and the chemical potential $\mu$ is assumed to be zero.
This expression is valid in the short-wavelength limit ($k\gg aH$), where the physical wavelength of the mode is much smaller than the Hubble radius.
Additionally, we assume the quasi-constant approximation for $m_{\rm eff}$, meaning that $m_{\rm eff}$ varies slowly compared to the time scale of the mode and can be treated as constant over short intervals.

According to the detailed calculation provided in Appendix~\ref{sec:calculation_of_noise_kernel}, we find the noise kernel in the form of
\begin{align}
    \mathcal{N}(p) = \mathcal{N}^{00}(p) + \mathcal{N}^{0\beta}(p) + \mathcal{N}^{\beta\beta}(p)\,.
\end{align}
Here, the free-free non-vanishing contribution is given by
\begin{align}
    \mathcal{N}^{00}(p)
    &= \frac{2\hbar^2}{5\pi} p^4 \left[\frac{1}{4}\left(1-\frac{4m^{2}_{\rm eff}}{p^2} \right) \right]^{5/2}\quad (4m^{2}_{\rm eff}<p^2)\,.
\end{align}

The free-thermal contribution identically vanishes,
\begin{align}
    \mathcal{N}^{0\beta}(p)
    &=0\,,
\end{align}
irrespective of the value of $p^2$.

Finally, we obtain the thermal-thermal contribution at high-temperature limit,
\begin{align}
    \tilde{N}(p)=  
    \begin{cases}
        & -\frac{\pi^2}{6p^2(p^2-p_0^2)^2\beta^2}\sqrt{1-\frac{4m^{2}_{\rm eff}}{p^2}}\Big( 11p^8 -46p^6p_0^2 + 82p^4 p_0^4 -44p^2 p_0^6 +4m^{2}_{\rm eff}(-5p^6+p^4p_0^2+2p^2p_0^4 + 2p_0^6 )\Big)\\
    & \qquad + \frac{\pi^2}{2p_0(-p^2+p_0^2)^{5/2}\beta^2}{\rm Arctanh}\left[ \frac{|{\bm p}|}{|p_0|}\sqrt{1-\frac{4m^{2}_{\rm eff}}{p^2}}\right] \Big( p^8 + 4p^6p_0^2 - 20p^4p_0^4+32p^2p_0^6 -16p_0^8\\
    &\qquad +16m^{4}_{\rm eff}(p^2-p_0^2)^2 -16m^{2}_{\rm eff}(p^2-p_0^2)^2(p^4-p^2p_0^2 + p_0^4)
    \Big)\,, \qquad\qquad(4m^{2}_{\rm eff}<p^2)\,,\\
    &0\,, \qquad\qquad(0<p^2<4m^{2}_{\rm eff})\,,\\
    & \frac{4\pi^2p^4}{3 (-p^2 +p_0^2)^{5/2}\beta^5}\,,
    \quad(p^2<0)\,,
    \end{cases}
\end{align}
at high temperature.

Hence, the reduced density matrix becomes
\begin{align}
    \rho_{r}[h^{+}_{f}, h^{-}_{f}, \eta_f] &=\int dh^{c}_{i} \int dh^{\Delta}_{i} \int_{h^{c}_{\bm p}(\eta_i)=h^{c}_{i}(\bm{p})}^{h^{c}_{\bm{p}}(\eta_f)=h^{c}_{f}(\bm{p})} D h^{c} \int_{h^{\Delta}_{\bm{p}}(\eta_i)=h^{\Delta}_{i}(\bm{p})}^{h^{\Delta}_{\bm{p}}(\eta_f)=h^{\Delta}_{f}(\bm{p})} Dh^{\Delta} \rho_{s}[h^{c}_{i}, h^{\Delta}_{i}, \eta_i]\\
    &\quad \times\exp\left[\frac{i}{\hbar} (S_0[h^{+}]-S_0[h^{-}]) + \int d^3 p \left( -\frac{1}{2\hbar^2} \int_{\eta_{i}}^{\eta_f} d\eta_1 \int_{\eta_{i}}^{\eta_f} d\eta_2\ h^{\Delta}_{\bm{p}}(\eta_1) h^{\Delta}_{-\bm{p}}(\eta_2) \mathcal{N}(\eta_2-\eta_1, {\bm p}) \right.\right.\\
    &\left.\left.-\frac{i}{2\hbar^2} \int_{\eta_{i}}^{\eta_f} d\eta_1 \int_{\eta_{i}}^{\eta_f} d\eta_2\ \theta(\eta_2-\eta_1)h^{c}_{\bm{p}}(\eta_1) h^{\Delta}_{-\bm{p}}(\eta_2) \mathcal{D}(\eta_2-\eta_1, {\bm p})\right) \right]\,,
\end{align}
in spatial momentum space.
Here, the dissipative kernel is given by
\begin{align}
    \mathcal{D}(\eta_2-\eta_1, {\bm p}) = 2 \int d^{3}k \left[ \bm{k}^2 -\frac{1}{\bm{p}^2}(\bm{p}\cdot\bm{k}) ^2\right]^2 {\rm Im}[G_{F}(\eta_2-\eta_1, {\bm k})G_{F}(\eta_2-\eta_1, -{\bm p}-{\bm k})]\,.
\end{align}
The time derivative of the reduced density matrix gives a master equation in the matrix element form
\begin{align}
 \frac{\partial}{\partial \eta_f}\rho_{r}[h^{+}_{f}, h^{-}_{f}, \eta_f] &= -\frac{i}{\hbar}\left[H_0\left(h^{+}_{f}({\bm p}), -i\hbar\frac{\delta}{\delta h^{+}_{f}({\bm p})}\right)-H_0\left(h^{-}_{f}({\bm p}), -i\hbar\frac{\delta}{\delta h^{-}_{f}({\bm p})}\right)\right] \rho_{r}[h^{+}_{f}, h^{-}_{f}, \eta_f] \\
 &-\frac{1}{\hbar^2} \int d^3 p\   h^{\Delta}_{f}(-{\bm p}) \int_{\eta_{i}}^{\eta_f} d\eta' \left[   \mathcal{N}(\eta_f-\eta', {\bm p}) \bar{h}^{\Delta}_{f}(\eta', {\bm p}) + \frac{i}{2} \mathcal{D}(\eta_f-\eta', {\bm p}) \bar{h}^{c}_{f}(\eta', {\bm p}) \right]\,,
 \label{eq:master_eq1}
\end{align}
where $H_{0}$ is the free Hamiltonian for $h$ and we have defined
\footnote{
In the operator form, the master equation can be expressed as
\begin{align}
     \frac{\partial}{\partial \eta_f}\hat{\rho}_{r}(\eta_f) &= -\frac{i}{\hbar}[\hat{H}_0, \hat{\rho}_{r}(\eta_f)] - \frac{1}{\hbar^2} \int d^3p \int^{\eta_f-\eta_i}_{0} d\tau \left\{  \mathcal{N}(\tau, {\bm p}) [\hat{h}^{\dagger}_{{\bm p}}(\eta_f),[\hat{h}_{{\bm p}}(\eta_f-\tau), \hat{\rho}_{r}(\eta_f)]] + \frac{i}{2} \mathcal{D}(\tau, {\bm p})[\hat{h}^{\dagger}_{\bm p}(\eta_f), \{\hat{h}_{\bm p}(\eta_f-\tau), \hat{\rho}_{r}(\eta_f)\}] \right\}\,.
\end{align}
However, we need to neglect the contribution of the influence functional to the full effective action in Eq.~\eqref{eq:master_eq2} to obtain this expression.
}
\begin{align}
    \bar{h}^{c, \Delta}_{f}(\eta', {\bm p}) := \int dh^{+}_{i} \int dh^{-}_{i} \int_{h^{+}_{\bm{p}}(\eta_i)=h^{+}_{i}(\bm{p})}^{h^{+}_{\bm{p}}(\eta_f)=h^{+}_{f}(\bm{p})} D h^{+} \int_{h^{-}_{\bm{p}}(\eta_i)=h^{-}_{i}(\bm{p})}^{h^{-}_{\bm{p}}(\eta_f)=h^{-}_{f}(\bm{p})} Dh^{-} \exp{\left[\frac{i}{\hbar}S_{\rm eff}[h^{+}, h^{-}]\right]} \rho_{h}[h^{+}_{i}, h^{-}_{i}, \eta_i] h^{c,\Delta}_{\bm{p}}(\eta')\,.
    \label{eq:master_eq2}
\end{align}

\section{Decoherence functional}
\label{sec:decoherence_functional}
Now that we have
\begin{align}
    S_{0}[h^{+}]-S_{0}[h^{-}] = -\frac{1}{2\kappa}\int d\eta \int d^3p\, a(\eta)^2&\left[ -\dot{h}^{c}_{\bm p}(\eta)\dot{h}^{\Delta}_{-{\bm p}}(\eta) 
    +|{\bm k}|^2 {h}^{c}_{\bm p}(\eta){h}^{\Delta}_{-{\bm p}}(\eta) 
    \right]\,,
\end{align}
variation of the effective action with respect to $h^{c}_{-{\bm p}}(\eta)$ and $h^{\Delta}_{-{\bm p}}(\eta)$ gives the classical equations of motion for $h^{\Delta}_{{\bm p}}(\eta)$ and $h^{c}_{\bm p}(\eta)$,
\begin{align}
    \ddot{h}^{cl}_{\Delta, {\bm p}}(\eta) + 2\mathcal{H}(\eta) \dot{h}^{cl}_{\Delta, {\bm p}}(\eta) + |{\bm p}|^2 h^{cl}_{\Delta, {\bm p}}(\eta) + \frac{\kappa}{\hbar}\frac{1}{a(\eta)^{2}}\int^{\eta_f}_{\eta} d\eta' h^{cl}_{\Delta, {\bm p}}(\eta')\mathcal{D}(\eta'-\eta, {\bm p}) = 0\,,
    \label{eq:classical_eom_hDelta}
\end{align}
and
\begin{align}
    \ddot{h}^{cl}_{c, {\bm p}}(\eta) + 2\mathcal{H}(\eta) \dot{h}^{cl}_{c, {\bm p}}(\eta)  + |{\bm p}|^2 h^{cl}_{c, {\bm p}}(\eta) + \frac{\kappa}{\hbar}\frac{1}{a(\eta)^{2}}\int^{\eta}_{\eta_i} d\eta'\ h^{cl}_{c, {\bm p}}(\eta')\mathcal{D}(\eta-\eta', {\bm p}) = i\frac{\kappa}{\hbar}\frac{1}{a(\eta)^{2}}\int^{\eta_f}_{\eta_i} d\eta'\ h^{cl}_{\Delta, {\bm p}}(\eta')\mathcal{N}(\eta-\eta', {\bm p})\,,
    \label{eq:classical_eom_hc}
\end{align}
respectively. Here, we have defined $\mathcal{H} := (a'/a)(\eta)$.
The homogeneous part of Eq.~\eqref{eq:classical_eom_hc} corresponds to the Abraham-Lorentz-like equation for the graviton.

Since the effective action is quadratic, the path integral in the reduced density matrix can be performed exactly if we replace the histories in the effective action with the classical solution and by taking into account fluctuations around the classical paths,
\begin{align}
    S_{\rm eff}[h^{+}_{cl}, h^{-}_{cl}] = 
    &-\frac{1}{2\kappa}\int d^3p\, a(\eta')^2\left[ -\dot{h}^{cl}_{c,{\bm p}}(\eta'){h}^{cl}_{\Delta, -{\bm p}}(\eta') \right]\Big|_{\eta_i}^{\eta_f}\cr
    &
    -\frac{1}{2\kappa}\int^{\eta_f}_{\eta_i} d\eta' \int d^3p\, a(\eta')^2\left(
    \ddot{h}^{cl}_{c, {\bm p}}(\eta') 
    + 2\mathcal{H}(\eta')\dot{h}^{cl}_{c,{\bm p}}(\eta')
    +|{\bm k}|^2 {h}^{cl}_{c, {\bm p}}(\eta')\right) 
    {h}^{cl}_{\Delta, -{\bm p}}(\eta') \cr 
    & + \int_{\eta_{i}}^{\eta_f} d\eta_1 \int_{\eta_{i}}^{\eta_f} d\eta_2 \int d^3 p \left\{ \frac{i}{2\hbar}  h^{cl}_{\Delta, {\bm p}}(\eta_1) h^{cl}_{\Delta, -{\bm p}}(\eta_2) \mathcal{N}(\eta_2-\eta_1, {\bm p}) \right.\cr
    & \hspace{4cm}
      \left. - \frac{1}{2\hbar}h^{cl}_{c, {\bm p}}(\eta_1) h^{cl}_{\Delta, -{\bm p}}(\eta_2) \mathcal{D}(\eta_2-\eta_1, {\bm p}) \right\}\,,
\end{align}
where $h^{\rm cl}_c$ and $h^{\rm cl}_{\Delta}$ are the classical solutions.
It is difficult to find the solution for Eq.~\eqref{eq:classical_eom_hc} due to the presence of the coupling between $h_{\Delta}$.
However, if we approximate the classical solution $h^{cl}_c$ by the homogeneous solution $h^{H}_c$ satisfying
\begin{align}
    \ddot{h}^{H}_{c, {\bm p}}(\eta) + 2\mathcal{H}(\eta) \dot{h}^{H}_{c, {\bm p}}(\eta) + |{\bm p}|^2 h^{H}_{c, {\bm p}}(\eta) + \frac{\kappa}{\hbar}\frac{1}{a(\eta)^{2}}\int^{\eta_f}_{\eta_i} d\eta_1 h^{H}_{c, {\bm p}}(\eta_1)\mathcal{D}(\eta-\eta_1, {\bm p}) = 0\,,
\end{align}
we find
\begin{align}
        \rho_{r}[h^{c}_{f}, h^{\Delta}_{f}, \eta_f] = N \int dh^{c}_{i} \int dh^{\Delta}_{i} \exp\left[\frac{1}{2\kappa}\frac{i}{\hbar}\int d^3 p \left\{a(\eta_f)^2\dot{h}^{c}_{f}h^{\Delta}_{f} - a(\eta_i)^2\dot{h}^{c}_{i}h^{\Delta}_{i}\right\} -\Gamma[h^{\Delta}_{i}, h^{\Delta}_{f}, \eta_f] \right] \rho_{r}[h^{c}_{i}, h^{\Delta}_{i}, \eta_i]\,,
\end{align}
where $N$ is the normalization constant given by the result of the path integral with respect to the fluctuations around the classical path, and $\Gamma$ is the decoherence functional,
\begin{align}
    \Gamma[h^{\Delta}_{i}, h^{\Delta}_{f}, \eta_f] = \frac{1}{2\hbar^2}\int_{\eta_{i}}^{\eta_f} d\eta_1 \int_{\eta_{i}}^{\eta_f} d\eta_2 \int d^3 p\,    h^{cl}_{\Delta, {\bm p}}(\eta_1) h^{cl}_{\Delta, -{\bm p}}(\eta_2) \mathcal{N}(\eta_2-\eta_1, {\bm p})\,,
    \label{eq:decoherence_functional}
\end{align}
which characterizes the fundamental decoherence between two coherent components of the GW
\footnote{While a full description of decoherence for a general initial state requires integrating over all components arising from its coherent-state decomposition, the decoherence functional provides a fundamental building block for such an analysis by isolating the interference between a pair of components.}
.

\section{Decoherence function}
\label{sec:decoherence_function}
By explicitly specifying the classical solutions in the decoherence functional, we evaluate the decoherence function. 
Since the contribution from the radiation-dominated era where $a(\eta)\propto\eta$ is expected to dominate, we approximate the classical solutions by the mode functions in the radiation-dominated era,
\begin{align}
    h^{\rm cl}_{\pm}(\eta, {\bm p}) = \frac{1}{|{\bm p}|\eta}\left(C_{\bm p}^{\pm}e^{-i|{\bm p}|\eta} + C_{-\bm p}^{\pm *}e^{i|{\bm p}|\eta} \right)\,.
\end{align}
Here, we require $C_{-{\bm p}}^{\pm *} = -C_{{\bm p}}^{\pm}$ to avoid divergence in the limit $\eta\rightarrow0$.

Since we assume the observation of GWs localized in a finite region, we consider Gaussian wave packets
\begin{align}
    C^{\pm}_{{\bm p}} = A_{\pm}\frac{1}{(2\pi \sigma^2_{\pm})^{3/2}} \exp\left(-\frac{|{\bm p}-{\bm p}_{\pm}|^2}{2\sigma_{\pm}^2} \right)\,,
\end{align}
for $p_z >0$ and determine $C^{\pm}_{{\bm p}}$ for $p_z <0$ through the condition $C_{-{\bm p}}^{\pm *} = -C_{{\bm p}}^{\pm}$.
Hereafter, we set ${\bm p}_{\pm}={\bm p}_{\rm GW}$ and $\sigma_{\pm}=\sigma_{\rm GW}$ for simplicity.

Substituting the classical solution into Eq~\eqref{eq:decoherence_functional} and performing integration in terms of $\eta_1$ and $\eta_2$ under $\eta_f\rightarrow\infty$ , we find
\begin{align}
    \Gamma[h^{\Delta}_{i}, h^{\Delta}_{f}, \eta_f] = \frac{1}{2}\frac{(A^{\Delta})^2 }{(2\pi\sigma^{2}_{\rm GW})^3} \int_{-\infty}^{\infty}dp_0\int d^3\! p\, F(p_0, {\bm p}) \frac{1}{|{\bm p}|^2}\exp\left(-2\frac{|{\bm p}-{\bm p}_{\rm GW}|^2}{2\sigma_{\rm GW}^2} \right)\tilde{N}(p)\,,
\end{align}
where we define a function by
\begin{align}
    F(p_0, {\bm p}) &= \lim_{\eta_f\rightarrow\infty}\left| \int^{\eta_f}_{\eta_i}\frac{d\eta}{\eta}\left(e^{i(p_0-|{\bm p}|)\eta}-e^{i(p_0+|{\bm p}|)\eta} \right) \right|^2\\
    &=\left|{\rm Ei}\left(i(p_0+|{\bm p}|)\eta_i\right) - {\rm Ei}\left(i(p_0-|{\bm p}|)\eta_i\right) \right|^2 \,.
\end{align}
with the exponential integral function ${\rm Ei}(z):=-\int^{\infty}_{-z}dt e^{-t}/t$.
The function $F$ is step-like with respect to $p_0$, maintaining a value $\sim 1/(1+(|{\bm p}| \eta_i)^{2})$ below $|{\bm p}|$ and rapidly approaching zero above $|{\bm p}|$.
The function $F$ determines how different energy-momentum components contribute to decoherence. 
In Minkowski spacetime, energy conservation constrains interactions to on-shell processes, keeping $F$ peaked at $p^2 = 0$.
In an expanding universe, time dependence alters the mode functions, redistributing spectral weight in $p_0$.
For $p_0 > |{\bm p}|$, rapid phase oscillations cause cancellation, suppressing $F$.
For $p_0 < |{\bm p}|$, this cancellation is weaker, allowing contributions to accumulate coherently.
This broadens the energy spectrum, making $p^2 < 0$ dominant in the decoherence process
\footnote{
If we assume a Minkowski background, the mode functions reduce to standard sinusoidal waves, and the function $F$ becomes a delta function centered around $p_{\rm GW}$. The noise kernel vanishes when the on-shell condition for the graviton, $p^2=0$, is satisfied, reflecting the fact that the decay of a graviton into two massive scalar particles is kinematically forbidden.
The bubble diagram under consideration cannot be cut into a pair of physical processes, 
and thus via the optical theorem its amplitude does not have a real part. Consequently, in the Minkowski background, there is no contribution to quantum decoherence at the one-loop level. 
Instead, the leading contribution is expected to arise from two-loop processes, which are suppressed by a factor inversely proportional to the Planck mass.
}.
We choose the scale factor to be unity at the reheating time, $\eta=\eta_r$, where the universe transitions to the radiation dominated phase.
We set $\eta_i = \eta_r$ and evaluate the quantities $m_{\rm eff},\ \beta,\ {\bm p}_{\rm GW}$ and $\sigma_{\rm GW}$ at $\eta_i=\eta_r$ denoting them as $m_{\rm eff,\,r}$, $\beta_{r}$, ${\bm p}_{{{\rm GW},r}}$, and $\sigma_{{\rm GW},r}$, respectively.
The initial amplitude $A^{\Delta}$ is determined by assuming instantaneous reheating after inflation.
Here, we consider that $m_{\rm eff}$ is approximated by the value at the reheating time
because the dominant contribution comes from an earlier stage of the radiation dominated era. 
Since at that time most of particles can be thought to be relativistic, we set $m_{{\rm eff},r}=0$.

Performing ${\bm p}$ integration by approximating
\begin{align}
    \frac{1}{(\pi\sigma^{2}_{{\rm GW},r})^{3/2}}\exp\left[-\frac{|{\bm p}-{\bm p}_{{\rm GW},r}|^2}{\sigma^2_{{\rm GW}, r}} \right]\simeq \delta^{(3)}({\bm p}-{\bm p}_{{\rm GW},r})\,,
\end{align}
we find
\begin{align}
    \Gamma[h^{\Delta}_{i}, h^{\Delta}_{f}, \eta_f] \approx \frac{(A^{\Delta})^2}{16(\pi\sigma^2_{{\rm GW},r})^{3/2}}   \int_{-\infty}^{\infty}dp_0\, \frac{1}{|{\bm p}_{{\rm GW}, r}|^2} F\left(p_0, {\bm p}_{{\rm GW},r}\right) N\left(p_0, {\bm p}_{{\rm GW},r}\right)\,.
    \label{eq:decoherence_function}
\end{align}
Hereafter, we set $\sigma_{{\rm GW},r} = p_{{\rm GW},r}$ as the decoherence function is simply scaled with $\sigma_{{\rm GW},r}^{-3}$. 
Since probably the number of oscillations should be at least $O(1)$ for the detection of GWs, this choice of $\sigma_{{\rm GW},r}$ would give the minimum value of $\Gamma$. 
The dominance of $p^2 < 0$ in $F$ implies that the noise kernel $N(p)$ is evaluated primarily in the spacelike region.
This contribution arises from thermal scatterings, where energy conservation is not exact due to the expansion of the universe.
While such processes are forbidden in Minkowski spacetime, the time-dependent background allows energy shifts, effectively opening new scattering channels.
Since these involve frequent interactions with thermal particles, the decoherence effect is significantly enhanced.
In this regime where $p^2<0$, the noise kernel can be estimated as
\begin{align}
    N\sim \frac{1}{p_{{\rm GW}, r} \beta_r^{5}}\,. 
\end{align}
The appearance of $\beta_r^{-5}$ can be understood by counting the dependences of each factor which appears in the loop integral: 
two number densities in Fourier space $n_{B}^2$, the loop integral $d^3k$, the two propagators and the contracted  
derivative indices give the factors $1/\beta^2$, $1/\beta^3$, $\beta^2$ and $1/\beta^2$, respectively, after the integral 
over the 0th component of the loop momentum. 
One may think that the last one may give a factor $1/\beta^4$, but we should recall that 
the contracted internal momentum is not $\propto \beta^{-2}$ but $\propto \beta^0$
\footnote{
For $p^2 > 4m^2$, $N(p)$ receives contributions from on-shell thermal scatterings.
These are standard kinematically allowed processes enhanced by the Bose-Einstein factor $(1 + n_B)$, leading to $N(p) \sim n_B^2 \propto T^2$.
However, at high temperatures, the number density of high-energy thermal particles is exponentially suppressed, making this contribution subdominant.
In contrast, the $p^2 < 0$ contribution, arising from a broader phase space integral, scales as $N(p) \propto T^5$ and dominates the decoherence process.
}. 
Using this estimate of $N$ and $F\sim \theta(p_{\rm GW}-|p_0|)/(1+(p_{\rm GW, r}/H_r)^{2})$, the order of magnitude of the decoherence functional can be estimated as
\begin{align}
    \Gamma \sim (A^{\Delta})^2 \left(\frac{H_r}{p_{{\rm GW},r}}\right)^{2} \left(\frac{T_r}{p_{{\rm GW},r}}\right)^{5}\,.
    \label{eq:Gamma_estimate}
\end{align}

\subsection{Model independent constraints}

We start with the analysis that can be done without specifying the details of the cosmic expansion model after inflation.
Here, we compute the amplitude threshold below which $\Gamma < 1$ is achieved for a given present-day GW frequency $f_{\rm GW, obs}$ and a reheating temperature $T_r$. 
We present the threshold GW amplitude in terms of $\Omega_{\rm GW}$, 
which represents the energy density of stochastic GWs $\rho_{\rm GW}$ per frequency bin,
\begin{align}
    \Omega_{\rm GW}:=\frac{1}{\rho_c}\frac{d\rho_{\rm GW}}{d \log{f}}\,,
\end{align}
normalized by the critical density $\rho_c = 3M_{\rm pl}^2H_0^2$.
$\Omega_{\rm GW}$ can be expressed using the present-day characteristic Fourier amplitude of GW signal $\tilde{h}_\cc(f)$ as
\begin{align}
    \Omega_{\rm GW}(f) = \frac{2\pi^2}{3H_0^{2}}f^2|\tilde{h}_\cc(f)|^2\,.
    \label{eq:def_Omega_GW}
\end{align}
Using $a_{\rm eq}/a_0\sim3\times10^{-4}$, we find $p_{\rm GW, 0}/H_0 \sim 0.02 (a_0/a_{*})$.
Here, the label $*$ represents the value at the horizon-crossing time assuming that the radiation-dominated phase lasts from a sufficiently early time, and the label ${\rm eq}$ refers to the value at the time of transition from radiation-dominated phase to the matter-dominated phase.
The relation $A^{\Delta}_0/A^{\Delta}\sim a_{*}/a_0$ connects the present-day critical amplitude $A^{\Delta}_{0, {\rm crit}}$, at which $\Gamma=1$ is achieved, to $A^{\Delta}_{{\rm crit}}$ as
\begin{align}
    (A^{\Delta}_{0, {\rm crit}})^2 =
        3\times10^{-4}\left(\frac{p_{\rm GW, 0}}{H_0}  \right)^{-2}(A^{\Delta}_{{\rm crit}})^2\,.
\end{align}
Eq.~\eqref{eq:decoherence_function} determines the critical amplitude $A^{\Delta}_{{\rm crit}}$ that gives $\Gamma =1 $.  
By substituting the above expression for $A^{\Delta}_{0, {\rm crit}}$ into $|\tilde{h}_\cc(f)|$ in Eq.~\eqref{eq:def_Omega_GW}, we find an amplitude threshold $\Omega_{\rm GW, max}$, below which quantum decoherence of GWs is inefficient.
By estimating $A^{\Delta}_{{\rm crit}}$ through Eq.~\eqref{eq:Gamma_estimate}, we can roughly evaluate $\Omega_{\rm GW, max}$ as 
\begin{align}
        \Omega_{\rm GW, max} \sim
        10^{-3}\left(\frac{f_{\rm GW, obs}}{5.6\times10^{11}\ {\rm Hz}}  \right)^{7}\left(\frac{T_r}{M_{\rm pl}}  \right)^{-2}\,.
\end{align}

\begin{figure}
    \centering
    \includegraphics[width=5.8in]{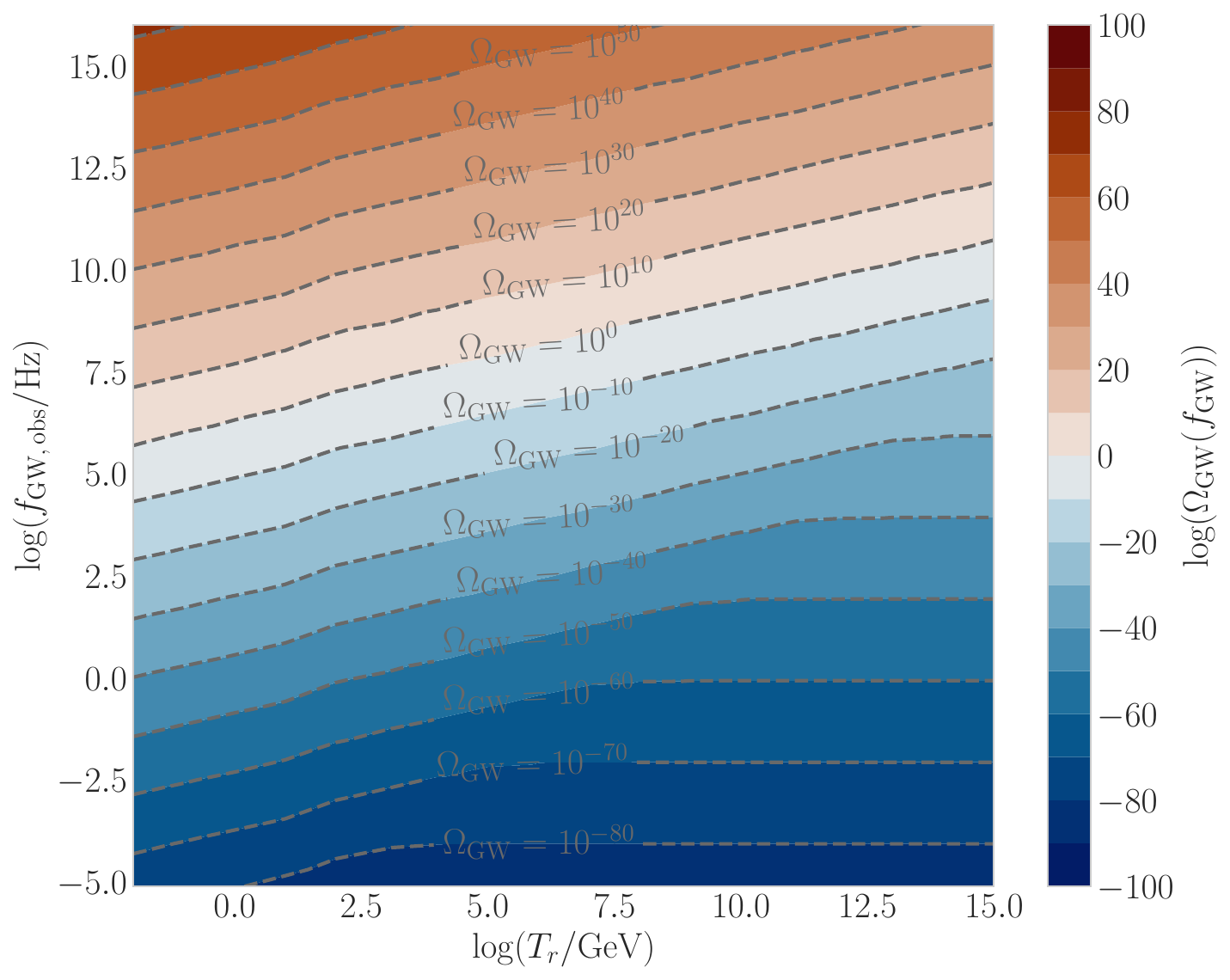}
    \caption{The maximum $\Omega_{\rm GW}$, below which quantum coherence can retain, is shown as a density plot. 
    The horizontal axis represents the reheating temperature $\log_{10}(T_{r}/{\rm GeV})$ and the vertical axis represents the observed GW frequency $\log_{10}(f_{{\rm GW,obs}}/{\rm Hz})$.
    We set $A^{\Delta}$ such that $\Gamma = 1$ for the given $T_r$ and $f_{\rm GW,obs}$.
    The other parameters are set by
    $\sigma_{{\rm GW},r} = p_{{\rm GW},r}$ and $m_{\rm eff} = 0\ {\rm GeV}$. 
    }
    \label{fig:OmegaGW_Tr_vs_fGW}
\end{figure}

Figure~\ref{fig:OmegaGW_Tr_vs_fGW} shows $\Omega_{\rm GW, max}$ as a function of the reheating temperature $T_r$ and the observed GW frequency $f_{\rm GW, obs}$, obtained by substituting $A^{\Delta}_{ {\rm crit}}$ derived from the numerical integration of Eq.~\eqref{eq:decoherence_function}.
The quantum decoherence effect can be negligible when we 
discuss the quantum coherence with an amplitude less than $\Omega_{\rm GW, max}$.
We should note that, the fact that the observed $\Omega_{\rm GW}$ of the stochastic GW background exceeds $\Omega_{\rm GW, max}$ does not mean that quantum coherence cannot be observed. 
Even if the total $\Omega_{\rm GW}$ is larger than the detection threshold, quantum coherence may still be detectable if the detector is sensitive enough to probe fine  structures associated with quantum properties.
The higher the reheating temperature and the lower the GW frequency, the stronger the effect of quantum decoherence, allowing only quantum coherence to be preserved within fine quantum fluctuations. 

\subsection{Model dependent constraints}
\begin{figure}
    \centering
    \includegraphics[width=5.8in]{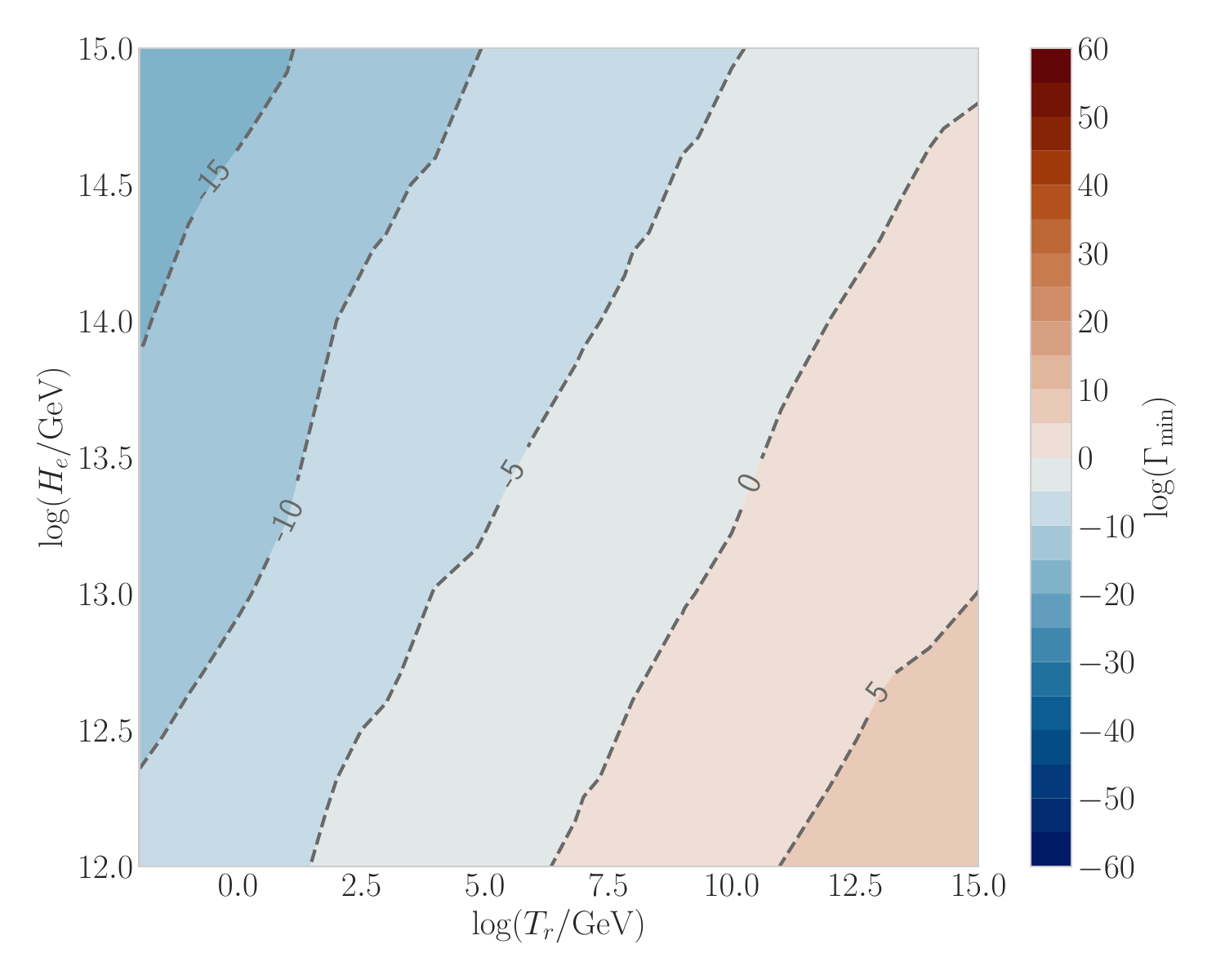}
    \caption{
    For the scenario in which the matter-dominant phase precedes the radiation dominant phase, the minimum value of the decoherence function $\Gamma_{\rm min}$ for the GWs with $p_{\rm GW}/H\simeq 1$ at the end of inflation is shown as a density plot. 
    The horizontal axis represents the reheating temperature $\log_{10}(T_{r}/{\rm GeV})$, and the vertical axis represents $\log_{10}(H_{e}/{\rm GeV})$.
    The color scale indicates $\log_{10}(\Gamma_{\rm min})$. 
    Parameters are set as $\sigma_{{\rm GW},r} = p_{{\rm GW},r}$, $A^{\Delta}_e = H_e/M_{\rm pl}$, and $m_{{\rm eff},r} = 0\ {\rm GeV}$. 
    {A lower temperature at the end of reheating and a larger Hubble parameter at the end of inflation suppress the effect of quantum decoherence of the GWs.}
}
    \label{fig:Tr_vs_He_MD}
\end{figure}

Here, we discuss in which case the original quantum state generated during inflation can remain effectively unaffected by the later propagation through the reheated universe. 
First, we consider the cosmological scenario in which the matter dominant phase follows after inflation, preceding the radiation dominant phase. 
We focus on the modes that are outside the horizon at the end of inflation and re-enter the horizon during the matter dominant phase. 
During the matter dominant phase, the amplitude of GWs decays by a factor of $(p_{{\rm GW},r}/H_r)^{-2}$.
On the other hand, if we assume that the instantaneous reheating after inflation, the amplitude decay would be $(p_{{\rm GW},r}/H_r)^{-1}$.
Therefore, for the modes that cross the horizon during the matter dominant phase, the GW amplitude should be reduced by the factor $(p_{{\rm GW},r}/H_r)^{-1}$ relative to the instantaneous reheating case.  
Namely, we should set the amplitude of GWs to 
\begin{align}
    A^{\Delta}=A^{\Delta}_e\left(\frac{p_{{\rm GW},r}}{H_r}\right)^{-1}\,,
    \label{eq:amp_MD}
\end{align}
where $A_e^{\Delta}$ is the primordial amplitude 
when the mode is on superhorizon scales.
Here, the label $e$ represents the time when inflation ends.
As the minimum value of $\Gamma$ is achieved for the highest frequency mode, we first consider the mode that is just at the horizon scale at the end of inflation, 
{\it i.e.}, the mode with $p_{{\rm GW},e}\approx H_e$. 
For this mode, since the GW momentum decays in proportion to the inverse of scale factor, we have $p_{{\rm GW},r}\approx T_r^{4/3} H_e^{1/3} M_{\rm pl}^{-2/3}$ written in terms of $T_{r}$ and $H_{e}$.
Hence, the GW excitation generated during inflation can remain undecohered only when 
\begin{align}
    \Gamma_{\rm min}\sim (A^{\Delta}_e)^2\left(\frac{T_{r}}{M_{\rm pl}}\right)\left(\frac{M_{\rm pl}}{H_{e}}\right)^{3}\sim 
    \left(\frac{T_{r}}{M_{\rm pl}}\right)\left(\frac{M_{\rm pl}}{H_{e}}\right)
    \,,
\end{align}
is smaller than unity. 
Figure~\ref{fig:Tr_vs_He_MD} shows this minimum value of the decoherence function $\Gamma_{\rm min}$ calculated by numerical integration of Eq.~\eqref{eq:decoherence_function} as a function of $T_r$ and $H_e$ for the GWs with $p_{\rm GW}/H\simeq 1$ at the end of inflation. 
{Here, we set $A^{\Delta}_{e}$ to the amplitude of the inflationary GWs, $H_e/M_{\rm pl}$, to discuss decoherence at the level of original GW amplitudes.}
In this case, a lower reheating temperature and a larger energy density at the end of inflation suppress the decoherence effect. 
Therefore, considering the standard history of the universe, in which the matter-dominated phase follows inflation before reheating, GWs are strongly decohered in parameter regions with a high reheating temperature, leading to the loss of quantum information when we discuss the quantum coherence at the level of the original amplitude generated by inflation.

\begin{figure}
    \centering
    \includegraphics[width=5.8in]{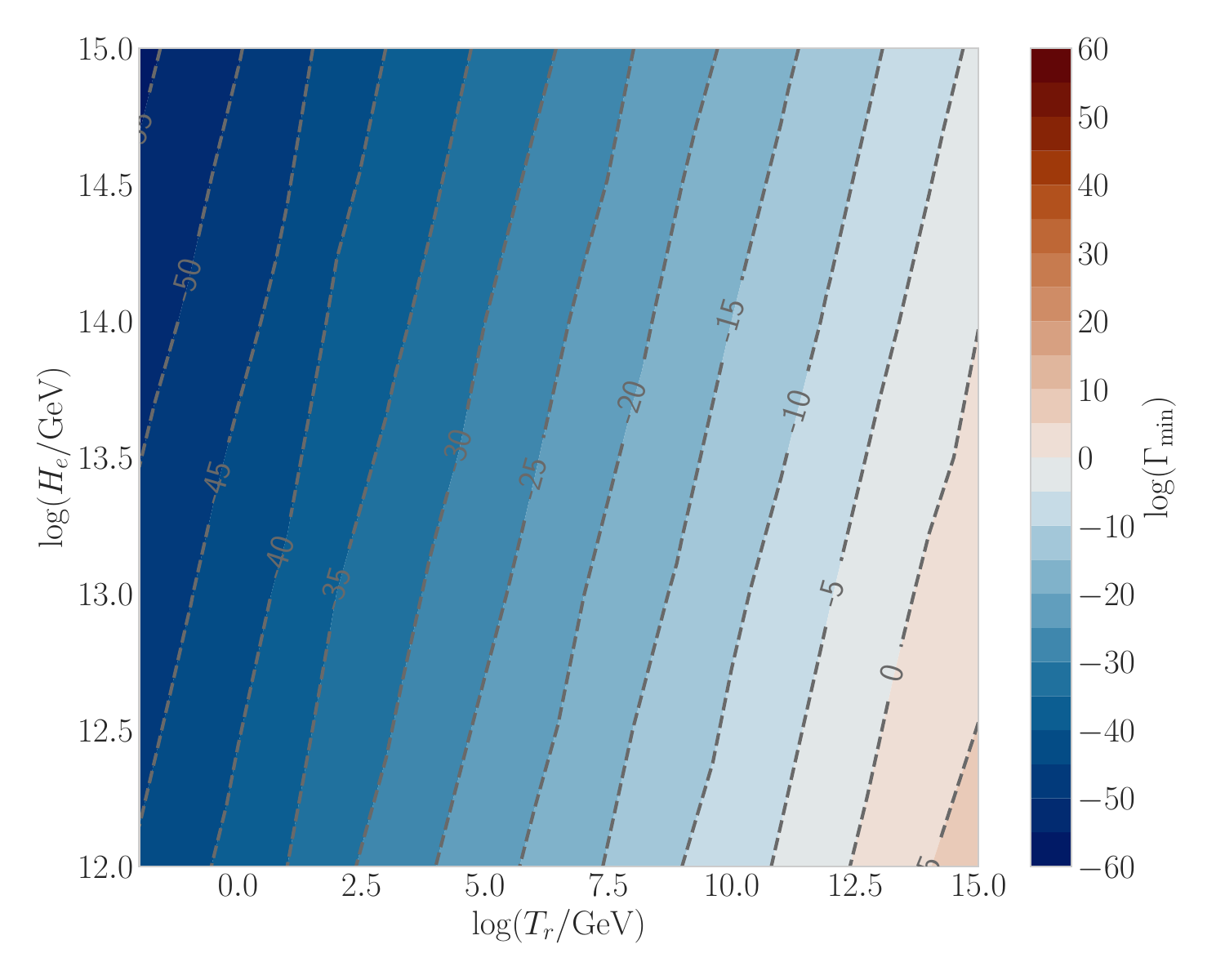}
    \caption{
    For the model with the kinetic-dominated phase, the minimum value of the decoherence function $\Gamma_{\rm min}$ for the GWs with $p_{\rm GW}/H\simeq1$ at the end of inflation is shown as a density plot. 
    The horizontal axis represents the reheating temperature $\log_{10}(T_{r}/{\rm GeV})$, and the vertical axis represents $\log_{10}(H_{e}/{\rm GeV})$.
    The color scale indicates $\log_{10}(\Gamma_{\rm min})$. 
    Parameters are set to $\sigma_{\rm GW} = p^{r}_{\rm GW}$, $A^{\Delta}_e = H_e/M_{\rm pl}$, and $m_{\rm eff,r} = 0\ {\rm GeV}$. 
    A lower reheating temperature and a larger energy density at the end of inflation more prominently suppress the decoherence effect compared with the model with the matter dominant phase.
}
    \label{fig:Tr_vs_He_KD}
\end{figure}

We now consider the case in which the kinetic term dominance phase follows after inflation. 
In this case the amplitude of GWs decays by $(p_{{\rm GW},r}/H_r)^{-1/2}$. 
Therefore, as in the preceding case, the amplitude of GWs is calculated to be enhanced by the factor $(p_{{\rm GW},r}/H_r)^{1/2}$ in this case.
Namely, we have 
\begin{align}
A^{\Delta}=A^{\Delta}_e \left(\frac{p_{{\rm GW},r}}{H_r}\right)^{1/2}\,. 
\label{eq:amp_KD}
\end{align}
As is known well, the amplitude of GWs in this scenario becomes larger at high frequencies compared with the standard scenario, which enhances the possibility of detecting primordial GWs. 
For the mode just at the horizon scale at the end of inflation, 
we have $p_{{\rm GW},r}= T_r^{2/3} H_e^{2/3} M_{\rm pl}^{-1/3}$.
Hence, we obtain
\begin{align}
    \Gamma_{\rm min}\sim (A^{\Delta}_e)^2\left(\frac{T_r}{M_{\rm pl}}\right)^{3}\left(\frac{M_{\rm pl}}{H_e}\right)^{4}
    \sim \left(\frac{T_r}{M_{\rm pl}}\right)^{3}\left(\frac{M_{\rm pl}}{H_e}\right)^{2}\,. 
\end{align}
Figure~\ref{fig:Tr_vs_He_KD} shows the minimum value of the decoherence function calculated by numerical integration of Eq.~\eqref{eq:decoherence_function} as a function of $T_r$ and $H_e$ for the GWs with $p_{\rm GW}/H\approx 1$ at the end of inflation.
Here, we set $A^{\Delta}_{e}$ to the amplitude of the inflationary GWs, $H_e/M_{\rm pl}$, to discuss decoherence at the level of original GW amplitudes.
As the power-law dependence on $T_r$ and $H_e$ changes, a lower temperature at the end of reheating and a larger Hubble parameter at the end of inflation more prominently suppress the decoherence effect. 
Consequently, the parameter region where the decoherence effect is negligible $\Gamma<1$ for the quantum state generated during inflation at the original amplitude level becomes wider.

\begin{figure}
    \centering
    \includegraphics[width=5.8in]{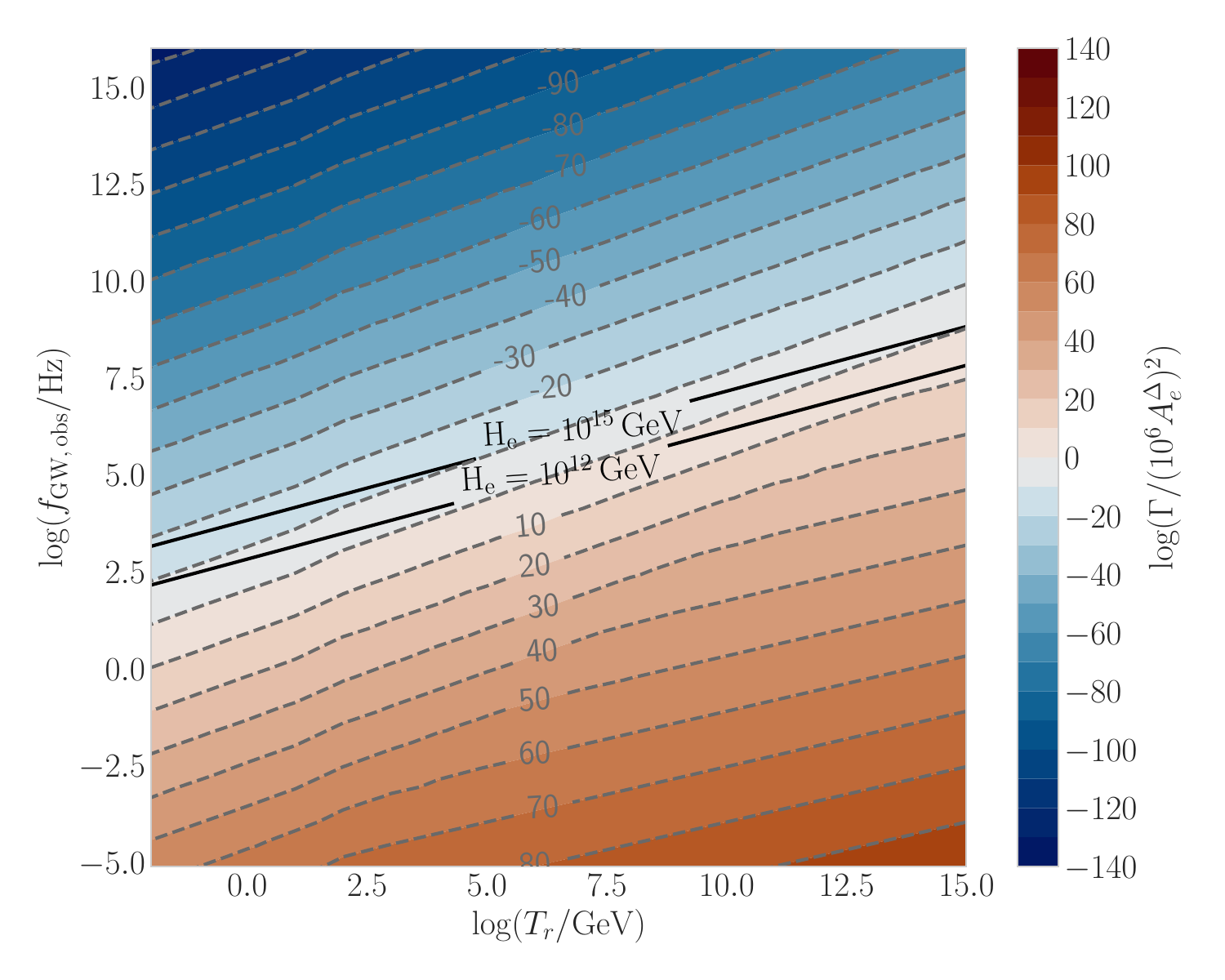}
    \caption{
    For model with the matter dominant phase, the scaled decoherence function $\Gamma/(10^{6}A^{\Delta}_{e})^2$ is shown as a density plot. 
    The horizontal axis represents the temperature $\log_{10}(T_{r}/{\rm GeV})$ and the vertical axis represents the observed GW frequency $\log_{10}(f_{{\rm GW,obs}}/{\rm Hz})$.
    The color scale indicates $\log_{10}(\Gamma/(10^{6}A^{\Delta}_{e})^2)$. 
    The dashed contours represent the frequencies of the observed GWs that are at the horizon scale at the end of inflation for several values of $H_e$.
    Parameters are set to $\sigma_{{\rm GW},r} = p_{{\rm GW},r}$ and $m_{\rm eff} = 0\ {\rm GeV}$. 
}
    \label{fig:Tr_vs_fGW_MD}
\end{figure}

\begin{figure}
    \centering
    \includegraphics[width=5.8in]{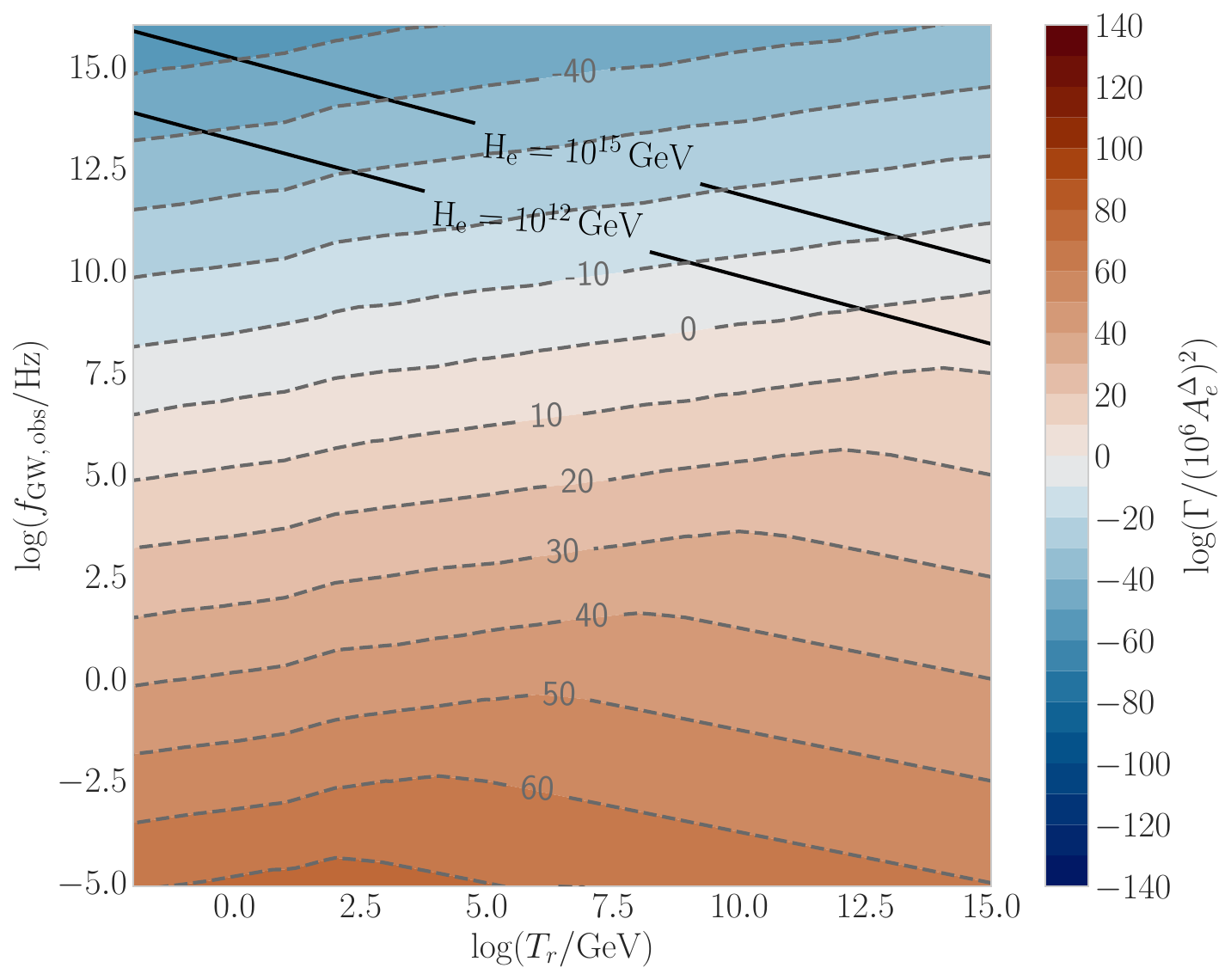}
    \caption{
    Same figure as Fig.~\ref{fig:Tr_vs_fGW_MD} but for the model with the kinetic dominant phase.
}
    \label{fig:Tr_vs_fGW_KD}
\end{figure}

Next, we investigate the dependence of the decoherence function on the observed GW frequency, assuming the GW amplitude generated during inflation.
By substituting Eqs.~\eqref{eq:amp_MD} and~\eqref{eq:amp_KD}
into Eq.~\eqref{eq:Gamma_estimate}, we estimate the decoherence function as the observed GW frequencies $f_{{\rm GW, obs}}$ as
\begin{align}
    \Gamma \sim
        (A^{\Delta}_{e})^2\left(\frac{T_{r}}{M_{\rm pl}}\right)^4 \left(\frac{f_{\rm GW, obs}}{5.6\times 10^{11}\ {\rm Hz}}\right)^{-9}\,,
\end{align}
for the model with the matter dominant phase and
\begin{align}
    \Gamma \sim(A^{\Delta}_{e})^2\left(\frac{T_{r}}{M_{\rm pl}}\right)\left(\frac{f_{\rm GW, obs}}{5.6\times 10^{11}\ {\rm Hz}}\right)^{-6}\,,
\end{align}
for the model with the kinetic dominant phase. 
Here, we have used the relation $p_{{\rm GW}, obs} = (T_0/T_r)p_{\rm GW, r}$.
Figure~\ref{fig:Tr_vs_fGW_MD} and Figure~\ref{fig:Tr_vs_fGW_KD} show the decoherence function calculated by numerical integration of Eq.~\eqref{eq:decoherence_function} as a function of the temperature $T_r$ and the observed GW frequency $f_{\rm GW, obs}$ for the respective models.
For several values of $H_e$, we also plot the lines corresponding to the frequencies of the observed GWs that enter the horizon scale exactly at the end of inflation.
Both figures illustrate, as a general tendency, a lower GW frequency further makes the system decohere more easily for a fixed reheating temperature.
{Additionally, in both scenarios, the system becomes less prone to decohere for a fixed GW frequency as the reheating temperature increases. 
This tendency is more pronounced in the model with the matter dominant phase.}
The difference in the dependence on the temperature arises from the different evolution of $A^{\Delta}$ during reheating.
{As shown in the figures, in the standard cosmological scenario, GWs generated by the inflation can remain effectively unaffected only in a limited parameter region. 
The observational frequency at which GWs can exhibit quantum nature strongly depends on the reheating temperature. 
Specifically, for $H_e \sim 10^{12}\ {\rm GeV}$, quantum state of GWs are modified by the effects of decoherence across all frequency bands when the reheating temperature is high. 
Only when the reheating temperature is sufficiently low, the effects of decoherence can be neglected in the $\sim 100\ {\rm Hz}$ band.  
In contrast, for $H_e \sim 10^{15}\ {\rm GeV}$, even at high reheating temperatures, the quantum state of GWs 
is not affected in the high-frequency band around $\sim 10^8\ {\rm Hz}$.
In the model with the kinetic dominant phase, the dependence on $T_r$ weakens, because of the enhanced amplitude of GWs at the present day. 
As a result, the observational frequency at which GWs can retain the original quantum state becomes higher than $\sim10^7\ {\rm Hz}$.}

\section{Conclusion}
\label{sec:conclusion}
We discussed quantum decoherence of gravitational waves (GWs) to assess whether GWs retain their quantum nature, using a simplified environmental model comprising a scalar field conformally coupled to gravity.
Tracing out the environmental degrees of freedom represented by the scalar field, we derived the reduced density matrix for the system field variable of a GW mode, which incorporated the effects of both dissipation and fluctuation, along with the master equation governing its time evolution.
A decoherence functional that quantifies the decay of the off-diagonal elements of the reduced density matrix was constructed using the noise kernel. 
Assuming Gaussian wave packets of GWs, we evaluated the decoherence function for two representative cosmological histories.

Our results indicate that quantum decoherence affects GWs more strongly at lower frequencies and a higher reheating temperature as a general trend.
As a model-independent result, we identified the amplitude threshold, below which GWs can retain their quantum coherence, in terms of the energy density of the observed GWs, $\Omega_{\rm GW, max}$.
We then investigated whether the initial quantum state generated during inflation can remain effectively unaffected by decoherence, assuming two representative models of cosmic expansion history. 
In the standard scenario, in which inflation is followed by the matter dominant phase before the radiation dominant phase begins, decoherence becomes stronger for a higher reheating temperature, at a fixed GW frequency (or momentum). 
When the energy density of the universe at the end of inflation is relatively low, the quantum state of GWs generated during inflation is modified by decoherence across all frequency bands for high reheating temperatures. 
In contrast, if the energy density at the end of inflation is relatively high, there exist regions of parameter space where the quantum state of GWs can retain their original quantum coherence, with a lower frequency bound ranging from $\sim 100\ {\rm Hz}$ to $\sim 10^8\ {\rm Hz}$ depending on the reheating temperature. 
In a model where the kinetic dominant phase precedes the radiation dominant phase, the dependence on the reheating temperature weakens, because the amplification of GWs during the kinetic dominant phase is larger for a lower reheating temperature.  
As a result, GWs can maintain the original quantum coherence set during inflation only for frequencies above $\sim 10^7\ {\rm Hz}$.

Drawing an analogy with quantum optics, detecting the quantum nature of GWs that cannot be explained by classical GWs would be essential to unambiguously claim the existence of gravitons as an experimentally proven fact.
Whether GWs can maintain their quantum nature depends both on the cosmological expansion history and the properties of the generated GWs
\footnote{
Several studies have discussed the decoherence of primordial tensor perturbations during inflation. 
These studies focus on two decoherence channels. 
One is the mechanism where short-wavelength modes act as the environment for the long-wavelength system~\cite{Gong:2019yyz, Burgess:2022nwu}. 
The other arises from interactions between the inflationary field and an external field that represents the environment~\cite{deKruijf:2024ufs}. 
In this work, we have explored a decoherence channel after inflation. 
A detailed investigation of the conditions under which each of these channels dominates remains beyond the scope of this paper and will be addressed in future research.
}.
If a mechanism exists that generates GWs in a non-trivial quantum state within the parameter region where quantum decoherence of GWs is negligible, and if humanity acquires a detector capable of probing the quantum nature of such GWs, it would, in principle, be possible to cleanly prove the quantumness of gravity. 
Furthermore, confirming the quantum nature of GWs could provide valuable insights into the physics of cosmic inflation, as the primordial tensor perturbations encode crucial information about the early universe.

\section*{Acknowledgements}
We would like to thank Danilo Artigas, Yanbei Chen, Akira Matsumura, Amaury Micheli, Yuta Michimura, Kimihiro Nomura, Hidetoshi Omiya, Jiro Soda, and Kazuhiro Yamamoto for useful comments.
H.T. is supported by the Hakubi project at Kyoto University and by Japan Society for the Promotion of Science (JSPS) KAKENHI Grant No. JP22K14037. 
T.T. is supported by JSPS KAKENHI Grant Nos. JP23H00110  24H00963, 24H01809 and JP20K03928.
\appendix
\section{Calculation of noise kernel}
\label{sec:calculation_of_noise_kernel}
We summarize the specific calculation of the noise kernel~\eqref{eq:noise_kernel},
\begin{align}
    \mathcal{N}(p) = 2 \int d^{4}k \left[ \bm{k}^2 -\frac{1}{\bm{p}^2}(\bm{p}\cdot\bm{k}) ^2\right]^2 {\rm Re}[G_{F}(k)G_{F}(-p-k)]\,,
    \label{eq:noise_kernel_appendix}
\end{align}
for the propagators~\eqref{eq:free_propagator} and~\eqref{eq:thermal_propagator},
\begin{align}
    G_{F}^0(k) = -i\hbar \frac{1}{-(k^{0})^2+\bm{k}^2+m^{2}-i\epsilon}\,,
\end{align}
and 
\begin{align}
    G_{F}^{\beta}(k) = 2\pi n_{B}(k^0) \delta(-(k^{0})^2+\bm{k}^2+m^{2})\,.
\end{align}
Note that we denote the effective mass $m_{\rm eff}$ as $m$ throughout the calculations below for notational convenience. 

The noise kernel can be decomposed,
\begin{align}
    \mathcal{N}(p) = \mathcal{N}^{00}(p) + \mathcal{N}^{0\beta}(p) + \mathcal{N}^{\beta\beta}(p)\,,
\end{align}
where the free-free part $\mathcal{N}^{00}(p)$ is given by
\begin{align}
   \mathcal{N}^{00}(p) = 2 \int d^{4}k \left[ \bm{k}^2 -\frac{1}{\bm{p}^2}(\bm{p}\cdot\bm{k}) ^2\right]^2 {\rm Re}[G^{0}_{F}(k)G^{0}_{F}(-p-k)]\,,
\end{align}
the free-thermal part $\mathcal{N}^{0\beta}(p)$ is given by
\begin{align}
   \mathcal{N}^{0\beta}(p) = 2 \int d^{4}k \left[ \bm{k}^2 -\frac{1}{\bm{p}^2}(\bm{p}\cdot\bm{k}) ^2\right]^2 {\rm Re}[G^{0}_{F}(k)G^{\beta}_{F}(-p-k)]\,,
\end{align}
and the thermal-thermal part $\mathcal{N}^{\beta\beta}(p)$ is given by
\begin{align}
   \mathcal{N}^{0\beta}(p) = 2 \int d^{4}k \left[ \bm{k}^2 -\frac{1}{\bm{p}^2}(\bm{p}\cdot\bm{k}) ^2\right]^2 {\rm Re}[G^{\beta}_{F}(k)G^{\beta}_{F}(-p-k)]\,.
\end{align}
In the following, we derive the free-free part, free-thermal part, and thermal-thermal part by performing 1-loop integrals. 

\subsection{Free-free part}
We evaluate the free-free part as
\begin{align}
    \mathcal{N}^{00}(p)&=2{\rm Re}\left[I_{1}^{00}(p)-\frac{2}{|\bm{p}|^2}I_{2}^{00}(p) + \frac{1}{|\bm{p}|^4}I_{3}^{00}(p)\right]\,,
\end{align}
where we have decomposed the contribution into three pieces
\begin{align}
    I_{1}^{00}(p)&=\int \frac{d^4 k}{(2\pi)^4} \bm{k}^4 G_{F}^{0}(k)G_{F}^{0}(-p-k) \,,\\
    I_{2}^{00}(p)&=\int \frac{d^4 k}{{(2\pi)^4}} \bm{k}^2 (\bm{p}\cdot\bm{k})^2 G_{F}^{0}(k)G_{F}^{0}(-p-k)\,,\\
    I_{3}^{00}(p)&=\int \frac{d^4 k }{{(2\pi)^4}}(\bm{p}\cdot\bm{k})^4 G_{F}^{0}(k)G_{F}^{0}(-p-k) \,.
\end{align}

According to dimensional regularization technique with the spacetime dimension expressed as $d = 4-2\eta$, Feynman parameter integral formula, and Wick's rotation, we find
\begin{align}
    I_{1}^{00}(p)&=\int \frac{d^D k}{(2\pi)^D} |\bm{k}|^4 \frac{-i\hbar}{(k+p)^2-m^2+i\epsilon} \frac{-i\hbar}{k^2-m^2+i\epsilon} \nonumber\\ 
    &= \frac{\Gamma(2)}{\Gamma(1)^2}\int^{1}_{0} dx \ \int \frac{d^D k'}{(2\pi)^D} |\bm{k}'-x\bm{p}|^4\frac{(-i\hbar)^2}{[k'^2 - \Delta^2 + i\epsilon]^2}\nonumber\\    
    &=\frac{\Gamma(2)}{\Gamma(1)^2}\int^{1}_{0} dx \ i(-1)^2\int \frac{d^D k_{E}}{(2\pi)^D} |\bm{k}_{E}-x\bm{p}|^4\frac{(-i\hbar)^2}{[|{k}_{E}|^2 + \Delta^2 - i\epsilon]^2}\nonumber\\
    &=\frac{\Gamma(2)}{\Gamma(1)^2} i(-1)^2 (-i\hbar)^2 \int^{1}_{0} dx \left[ J_1 + 4x^2 p_{i}p_{j}J^{ij}_{7} + x^4 |\bm{p}^4| J_9 -4x p_{i} J^{i}_3 +2x^2|\bm{p}^2|J_2 -4x^3 |\bm{p}|^2 p_{i} J^{i}_8 \right]\,,
\end{align}
\begin{align}
    I_{2}^{00}(p)&=\int \frac{d^D k}{{(2\pi)^D}} |\bm{k}|^2 (\bm{p}\cdot\bm{k})^2 \frac{-i\hbar}{(k+p)^2-m^2+i\epsilon} \frac{-i\hbar}{k^2-m^2+i\epsilon} \nonumber\\
    &=\frac{\Gamma(2)}{\Gamma(1)^2}\int^{1}_{0} dx \ i(-1)^2\int \frac{d^D k_{E}}{(2\pi)^D} |\bm{k}_{E}-x\bm{p}|^2 \left\{ (\bm{k}_E -x\bm{p})\cdot\bm{p}\right\}^2\frac{(-i\hbar)^2}{[|{k}_{E}|^2 + \Delta^2 - i\epsilon]^2}\nonumber\\
    &=\frac{\Gamma(2)}{\Gamma(1)^2} i(-1)^2 (-i\hbar)^2 \int^{1}_{0} dx \left[ p_i p_j J_{4}^{ij} -2x|{\bm p}|^2 p_{i} J_{3}^{i} +x^2 |{\bm p}^4|J_{2} -2x p_{i}p_{j}p_{l}J_{6}^{ijl} \right.\nonumber\\
    &\left.\ + 5x^2|{\bm p}^2|p_{i}p_{j}J_{7}^{ij} -4x^3|\bm{p}^4|p_{i}J_{8}^i + x^4 |{\bm p}|^6 J_{9} \right]\,,
\end{align}
\begin{align}
    I_{3}^{00}(p)&=\int \frac{d^D k}{{(2\pi)^D}}(\bm{k}\cdot\bm{p})^4 \frac{-i\hbar}{(k+p)^2-m^2+i\epsilon} \frac{-i\hbar}{k^2-m^2+i\epsilon} \nonumber\\
    &=\frac{\Gamma(2)}{\Gamma(1)^2}\int^{1}_{0} dx \ i(-1)^2\int \frac{d^D k_{E}}{(2\pi)^D} \left\{ (\bm{k}_E -x\bm{p})\cdot\bm{p}\right\}^4\frac{(-i\hbar)^2}{[|{k}_{E}|^2 + \Delta^2 - i\epsilon]^2}\nonumber\\
    &=\frac{\Gamma(2)}{\Gamma(1)^2} i(-1)^2 (-i\hbar)^2 \int^{1}_{0} dx \left[ p_{i}p_{j}p_{l}p_{m}J_{5}^{ijlm} + 6x^2|{\bm p}|^4 p_{i}p_{j}J_{7}^{ij} + x^4|{\bm p}|^8 J_{9} -4x|{\bm p}|^2p_{i}p_{j}p_{l}J_{6}^{ijl} -4x^3|{\bm p}|^6 p_{i}J_{8}^{i} \right]\,,
\end{align}
where we have defiend $k':=k+xp$, the Euclid momentum $k_{E} = (k_{E}^0, \bm{k}_{E})$, and $\Delta^2:=m^2-x(1-x)p^2$.
Here, the quantities $J_1$-$J_9$ are defined as follows: 
\begin{align}
    J_1 &= \int \frac{d^D k_{E}}{(2\pi)^D} \frac{|{\bm k}_{E}|^4}{[|{k}_{E}|^2 + \Delta^2 - i\epsilon]^2}\nonumber\\
    &=\frac{\Gamma(\frac{D}{2})\Gamma(\frac{3+D}{2})}{\Gamma(2+\frac{D}{2})\Gamma(\frac{-1+D}{2})}\int \frac{d^D k_{E}}{(2\pi)^D} \frac{|{k}_{E}|^4}{[|{k}_{E}|^2 + \Delta^2 - i\epsilon]^2}\nonumber\\
    &=\frac{\Gamma(\frac{D}{2})\Gamma(\frac{3+D}{2})}{\Gamma(2+\frac{D}{2})\Gamma(\frac{-1+D}{2})} \frac{D(D+2)}{2}\frac{1}{(4\pi)^{D/2}}\frac{\Gamma(2-D/2-1)}{\Gamma(2)}(\Delta^2 -i\epsilon)^{-2+D/2+2}\nonumber\\
    &= -\frac{15}{2}\left[ m^2+(p_0^2 -|{\bm p}|^2)x(x-1)\right]^2 \left(J_9 -\frac{1}{15}\right)\nonumber\\
    &= -\frac{15\left[ m^2+(p_0^2 -|{\bm p}|^2)x(x-1)\right]^2}{32\pi^2}\left[ \frac{1}{\eta} -\gamma +\log(4\pi) + 1 + \psi^{(0)}\left(\frac{3}{2}\right) - \psi^{(0)}\left(\frac{7}{2}\right) -\log\left[m^2 + (p_0^2 - |{\bm p}|^2) x(x-1) \right] \right]\,,
\end{align}
\begin{align}
    J_2 &= \int \frac{d^D k_{E}}{(2\pi)^D} \frac{|{\bm k}_{E}|^2}{[|{k}_{E}|^2 + \Delta^2 - i\epsilon]^2}\nonumber\\
    &=\frac{\Gamma(\frac{D}{2})\Gamma(\frac{1+D}{2})}{\Gamma(1+\frac{D}{2})\Gamma(\frac{-1+D}{2})}\int \frac{d^D k_{E}}{(2\pi)^D} \frac{|{k}_{E}|^2}{[|{k}_{E}|^2 + \Delta^2 - i\epsilon]^2}\nonumber\\
    &=\frac{\Gamma(\frac{D}{2})\Gamma(\frac{1+D}{2})}{\Gamma(1+\frac{D}{2})\Gamma(\frac{-1+D}{2})} \frac{D}{2}\frac{1}{(4\pi)^{D/2}}\frac{\Gamma(2-D/2-1)}{\Gamma(2)}(\Delta^2 -i\epsilon)^{-2+D/2+1}\nonumber\\
    &= -\frac{3}{2}\left[ m^2+(p_0^2 -|{\bm p}|^2)x(x-1)\right] \left(J_9 + \frac{1}{3} \right)\nonumber\\
    &= -\frac{3\left[ m^2+(p_0^2 -|{\bm p}|^2)x(x-1)\right]}{32\pi^2}\left[ \frac{1}{\eta} -\gamma +\log(4\pi) + 1 + \psi^{(0)}\left(\frac{3}{2}\right) - \psi^{(0)}\left(\frac{5}{2}\right) -\log\left[m^2 + (p_0^2 - |{\bm p}|^2) x(x-1) \right] \right]\,,
\end{align}
\begin{align}
    J_{3}^{i} &= \int \frac{d^D k_{E}}{(2\pi)^D} \frac{|{\bm k}_{E}|^2 k_{E}^{i}}{[|{k}_{E}|^2 + \Delta^2 - i\epsilon]^2}\nonumber\\
    &=0\,,
\end{align}
\begin{align}
    J_4^{ij} &= \int \frac{d^D k_{E}}{(2\pi)^D} \frac{|{\bm k}_{E}|^2 k_{E}^{i} k_{E}^{j}}{[|{k}_{E}|^2 + \Delta^2 - i\epsilon]^2}\nonumber\\
    &=\frac{1}{D-1}\delta_{ij}J_{1}\,,
\end{align}
\begin{align}
    J_5^{ijlm} &= \int \frac{d^D k_{E}}{(2\pi)^D} \frac{k_{E}^{i} k_{E}^{j} k_{E}^{l} k_{E}^{m}}{[|{k}_{E}|^2 + \Delta^2 - i\epsilon]^2}\nonumber\\
    &=\frac{1}{(D-1)^2+2(D-1)}(\delta_{ij}\delta_{lm}+\delta_{il}\delta_{jm}+\delta_{im}\delta_{jl})J_1\,,
\end{align}
\begin{align}
    J_6^{ijl} &= \int \frac{d^D k_{E}}{(2\pi)^D} \frac{k_{E}^{i} k_{E}^{j} k_{E}^{l} }{[|{k}_{E}|^2 + \Delta^2 - i\epsilon]^2}\nonumber\\
    &=0\,,
\end{align}
\begin{align}
    J_7^{ij} &= \int \frac{d^D k_{E}}{(2\pi)^D} \frac{k_{E}^{i} k_{E}^{j}}{[|{k}_{E}|^2 + \Delta^2 - i\epsilon]^2}\nonumber\\
    &= \frac{1}{D-1} \delta_{ij} J_{2}\,,
\end{align}
\begin{align}
    J_8^{i} &= \int \frac{d^D k_{E}}{(2\pi)^D} \frac{k_{E}^{i}}{[|{k}_{E}|^2 + \Delta^2 - i\epsilon]^2}\nonumber\\
    &=0\,,
\end{align}
\begin{align}
    J_9 &= \int \frac{d^D k_{E}}{(2\pi)^D} \frac{1}{[|{k}_{E}|^2 + \Delta^2 - i\epsilon]^2}\nonumber\\
    &= \frac{\Gamma(2-D/2)}{(4\pi)^{D/2}\Gamma(2)}(\Delta^2-i\epsilon)^{-2+D/2}\nonumber\\
    &= \frac{1}{16\pi^2}\left[ \frac{1}{\eta} - \gamma +\log(4\pi) - \log\left[m^2 + (p_0^2 - |{\bm p}|^2) x(x-1) \right]\right]\,,
\end{align}
where $\gamma$ is the Euler's constant and $\psi^{(n)}(z):= (d^n/dz^n)(\Gamma'(z)/\Gamma(z))$ with the Gamma function $\Gamma(z)$.

Hence, we find
\begin{align}
    I_{1}^{00}(p)&=\frac{\Gamma(2)}{\Gamma(1)^2} i(-1)^2 (-i\hbar)^2 \int^{1}_{0} dx \left[ J_1 + \frac{4}{D-1}x^2 |\bm{p}|^2J_{2} + x^4 |\bm{p}|^4 J_9  +2x^2|\bm{p}|^2J_2 \right]\nonumber\\
    &= -i\hbar^2 \int^{1}_{0} dx \left[ -\frac{5}{3}|{\bm p}|^2 x^2 \Delta^2 + \frac{1}{2}\Delta^4 + J_9\left( |{\bm p}|^4x^4 - 5|{\bm p}|^2 x^2 \Delta^2 -\frac{15}{2}\Delta^4 \right) \right]\,,
\end{align}
\begin{align}
    I_{2}^{00}(p)&=\frac{\Gamma(2)}{\Gamma(1)^2} i(-1)^2 (-i\hbar)^2 \int^{1}_{0} dx \left[ \frac{1}{D-1}|\bm{p}|^2 J_{1}  +x^2 |{\bm p}|^4J_{2}+ \frac{5}{D-1}x^2|{\bm p}|^4J_{2}  + x^4 |{\bm p}|^6 J_{9} \right]\nonumber\\
    &= -i\hbar^2 \int^{1}_{0} dx \left[ -\frac{4}{3}|{\bm p}|^4 x^2 \Delta^2 + \frac{1}{6}|{\bm p}|^2\Delta^4 + J_9\left( |{\bm p}|^6x^4 - 4|{\bm p}|^4 x^2 \Delta^2 -\frac{5}{2}|{\bm p}|^2\Delta^4 \right) \right]\,,
\end{align}
\begin{align}
    I_{3}^{00}(p)&=\frac{\Gamma(2)}{\Gamma(1)^2} i(-1)^2 (-i\hbar)^2 \int^{1}_{0} dx \left[ \frac{1}{(D-1)^2+2(D-1)}|\bm{p}|^4 J_{1} + \frac{6}{D-1}x^2|{\bm p}|^6 J_{2} + x^4|{\bm p}|^8 J_{9} \right]\nonumber\\
    &= -i\hbar^2 \int^{1}_{0} dx \left[ -|{\bm p}|^6 x^2 \Delta^2 + \frac{1}{30}|{\bm p}|^4\Delta^4 + J_9\left( |{\bm p}|^8 x^4 - 3|{\bm p}|^6 x^2 \Delta^2 -\frac{1}{2}|{\bm p}|^4\Delta^4 \right) \right]\,.
\end{align}

Finally, the free-free part becomes
\begin{align}
    \mathcal{N}^{00}(p)&=2{\rm Re}\left[I_{1}^{00}(p)-\frac{2}{|\bm{p}|^2}I_{2}^{00}(p) + \frac{1}{|\bm{p}|^4}I_{3}^{00}(p)\right]\nonumber\\
    &= {\rm Re}\left[ -\frac{2}{5}i\hbar^2\int^{1}_{0} dx \left[ \Delta^4 (1 - 15 J_9) \right] \right]\nonumber\\
    &= {\rm Re}\left[ -\frac{2}{5}i\hbar^2\int^{1}_{0} dx \left[ \Delta^4 \left\{1 - \frac{15}{16\pi^2}\left( \frac{1}{\eta} - \gamma +\log(4\pi) - \frac{1}{2}\log\Delta^2\right)\right\}\right] \right]\nonumber\\
    &= {\rm Re}\left[ -\frac{1}{2400\pi^2}i\hbar^2 \left[ 900\gamma m^4 + 960m^4(\pi^2-1) + (p_0^2 -|{\bm p}|^2)^2(-47+30\gamma +32\pi^2)\right.\right.\nonumber\\
    &\quad-10m^2 (p_0^2 -|{\bm p}|^2) (-41+30\gamma +32\pi^2) \nonumber \\
    &\quad- \frac{1}{\eta} 30\Big(30m^4-10m^2(p_0^2 -|{\bm p}|^2) + (p_0^2 -|{\bm p}|^2)^2\Big) + I \nonumber\\
    &\quad\left.\left.+ 30\Big(30m^4-10m^2(p_0^2 -|{\bm p}|^2) + (p_0^2 -|{\bm p}|^2)^2\Big) \log{\frac{m^2}{4\pi}}\right]\right]\nonumber\\
    &= \frac{2\hbar^2}{5\pi} p^4 C^{5/2}\quad (4m^2<p^2) \,,
\end{align}
where 
\begin{align}
    I = 
    \begin{cases}
        60\cdot 16C^2 p^4 2i \sqrt{C} \frac{1}{4}\left(-2\pi + i \log\left[ \frac{\sqrt{C}-\frac{1}{2}}{\sqrt{C}+\frac{1}{2}}\right]  \right)\quad (4m^2<p^2)\\
        60\cdot 16C^2 p^4 2\sqrt{-C} \arctan\left[{\frac{1}{2\sqrt{-C}}}\right]\quad (0<p^2<4m^2)\\
        60\cdot 16C^2 p^4 2i \sqrt{C} \frac{1}{4} 2i \log\left[ \frac{\sqrt{C}-\frac{1}{2}}{\sqrt{C}+\frac{1}{2}}\right] \quad (p^2<0) \,,\\
    \end{cases}
\end{align}
with 
\begin{align}
    C := \frac{1}{4}\left[ 1- \frac{4m^2}{p^2} \right]\,. 
\end{align}

\subsection{Free-thermal part}
We evaluate the free-thermal part as
\begin{align}
    \mathcal{N}^{0\beta}(p)&=4{\rm Re}\left[I_{1}^{0\beta}(p)-\frac{2}{|\bm{p}|^2}I_{2}^{0\beta}(p) + \frac{1}{|\bm{p}|^4}I_{3}^{0\beta}(p)\right]\,,
\end{align}
where we have decomposed the contribution into three pieces
\begin{align}
    I_{1}^{0\beta}(p)&=\int \frac{d^4 k}{{(2\pi)^4}} \bm{k}^4 G_{F}^{0}(k)G_{F}^{\beta}(-p-k)\,, \\
    I_{2}^{0\beta}(p)&=\int \frac{d^4 k}{{(2\pi)^4}} \bm{k}^2 (\bm{p}\cdot\bm{k})^2 G_{F}^{0}(k)G_{F}^{\beta}(-p-k)\,, \\
    I_{3}^{0\beta}(p)&=\int \frac{d^4 k}{{(2\pi)^4}} (\bm{p}\cdot\bm{k})^4 G_{F}^{0}(k)G_{F}^{\beta}(-p-k)\,.
\end{align}

On substituting the propagators, we can rewrite the pieces as
\begin{align}
    I_{1}^{0\beta}(p)&=\int \frac{d^4 k}{{(2\pi)^4}} |\bm{k}|^4 \frac{-i\hbar}{(p+k)^2-m^2+i\epsilon}\times 2\pi n_{B}(k^{0})\delta(k^2-m^2)\nonumber\\
    &=(-i\hbar)2\pi \int \frac{d^4 k_2}{{(2\pi)^4}} \bm{k}_{2}^4 \delta((p+k)^2-m^2) n_{B}(k^{0})\delta(k^2-m^2)\nonumber\\
    &=(-i\hbar)2\pi \frac{1}{{(2\pi)^4}} \int_{-\infty}^{\infty}dk_0 \int_{0}^{\infty}d|\bm{k}| |{\bm k}|^2 |\bm{k}|^4 n_B(k_0) \delta(k^2-m^2) \int_{-1}^{1}d\cos{\theta} \delta(p^2 +2p_0k_0-2|{\bm p}||{\bm k}|\cos{\theta}+k^2-m^2)\,,
\end{align}
\begin{align}
    I_{2}^{0\beta}(p)&=\int \frac{d^4 k}{{(2\pi)^4}} |\bm{k}|^2 (\bm{p}\cdot\bm{k})^2 \frac{-i\hbar}{(p+k)^2-m^2+i\epsilon}\times 2\pi n_{B}(k^{0})\delta(k^2-m^2)\nonumber\\
    &=(-i\hbar)2\pi \frac{1}{{(2\pi)^4}} \int_{-\infty}^{\infty}dk_0 \int_{0}^{\infty}d|\bm{k}| |{\bm k}|^2 |\bm{k}|^2 (\bm{p}\cdot\bm{k})^2 n_B(k_0) \delta(k^2-m^2) \nonumber\\ &\quad\times\int_{-1}^{1}d\cos{\theta} \delta(p^2 +2p_0k_0-2|{\bm p}||{\bm k}|\cos{\theta}+k^2-m^2)\,,
\end{align}
\begin{align}
    I_{3}^{0\beta}(p)&=\int \frac{d^4 k}{{(2\pi)^4}} (\bm{p}\cdot\bm{k})^4 \frac{-i\hbar}{(p+k)^2-m^2+i\epsilon}\times 2\pi n_{B}(k^{0})\delta(k^2-m^2)\nonumber\\
    &=(-i\hbar)2\pi \frac{1}{{(2\pi)^4}} \int_{-\infty}^{\infty}dk_0 \int_{0}^{\infty}d|\bm{k}| |{\bm k}|^2 (\bm{p}\cdot\bm{k})^4 n_B(k_0) \delta(k^2-m^2) \nonumber \\ &\quad\times\int_{-1}^{1}d\cos{\theta} \delta(p^2 +2p_0k_0-2|{\bm p}||{\bm k}|\cos{\theta}+k^2-m^2)\,.
\end{align}

By the existence of the delta functions, only $k$ satisfying the following condition can contribute the integral,
\begin{align}
    \frac{|p^2+2p_0k_0|}{2|{\bm p}||{\bm k}|}<1,\quad {\rm and}\quad k_0^2=|\bm{k}|^2+m^2\,.
\end{align}
This condition reduce to the conditions on $k$~\cite{Nishikawa:2003js},
\begin{align}
    {\rm A.}\quad p^2<0,\quad p_0k_0p^2>0,\quad {\rm and}\quad |{\bm k}|^2>|{\bm k}|^2_{+}\,,
\end{align}
\begin{align}
    {\rm B.}\quad &p^2>4m^2,\quad p_0k_0p^2<0,\quad {\rm and}\quad |{\bm k}|_{-}^2<|{\bm k}|^2<|{\bm k}|^2_{+}\,,\\
    &{\rm or}\quad p^2<0,\quad p_0k_0p^2<0,\quad {\rm and}\quad |{\bm k}|_{-}^2<|{\bm k}|^2<|{\bm k}|^2_{0}\,,
\end{align}
\begin{align}
    {\rm C.}\quad p^2<0,\quad p_0k_0p^2<0,\quad {\rm and}\quad |{\bm k}|^2>|{\bm k}|^2_{0}\,,
\end{align}
where
\begin{align}
    |{\bm k}|_{\pm}^2 = \frac{1}{4}\left\{ |{\bm p}| \pm \sqrt{1-\frac{4m^2}{p^2}}|p_0| \right\}^2\,,
\end{align}
giving
\begin{align}
    \omega_{\pm} = \frac{1}{2} \left| |p_0| \pm \sqrt{1-\frac{4m^2}{p^2}}|{\bm p}| \right|\,.
\end{align}
Note that the second condition corresponding to case B and the condition for case C can be combined taking into account $|{\bm k}|_{-}^2<|{\bm k}|^2$. Thus, the noise kernel is divided into three regions depending on the value of $p^2$ as
\begin{align}
I_{1}^{0\beta} =
\begin{cases}
& (-i\hbar)\frac{\pi}{|{\bf p}|}\int_{\omega_{-}}^{\omega_{+}}d\omega_k n_B(\omega_k)(\omega_k^2-m^2)^2\quad (4m^2<p^2)\\ \\
& 0\quad (0<p^2<4m)\\ \\
& (-i\hbar)\frac{\pi}{|{\bf p}|}\left(\int_{\omega_{-}}^{\infty} +\int_{\omega_{+}}^{\infty} \right)d\omega_k n_B(\omega_k)(\omega_k^2-m^2)^2\quad (p^2<0)\,,
\end{cases}
\end{align}
\begin{align}
I_{2}^{0\beta} =
\begin{cases}
& (-i\hbar)\frac{\pi}{2|{\bf p}|}\int_{\omega_{-}}^{\omega_{+}}d\omega_k (\omega_k^2-m^2)(p^2-2|p_0|\omega_k)^2n_B(\omega_k) \quad (4m^2<p^2)\\ \\
& 0\quad (0<p^2<4m^2)\\ \\
& (-i\hbar)\frac{\pi}{4|{\bf p}|}\left(\int_{\omega_{-}}^{\infty} d\omega_k (\omega_k^2-m^2)(p^2+|p_0|\omega_{k})^2n_B(\omega_k)+\int_{\omega_{+}}^{\infty} d\omega_k (\omega_k^2-m^2)(p^2-|p_0|\omega_{k})^2n_B(\omega_k) \right)\quad (p^2<0)\,,
\end{cases}
\end{align}
\begin{align}
I_{3}^{0\beta} =
\begin{cases}
& (-i\hbar)\frac{\pi}{2|{\bf p}|}\int_{\omega_{-}}^{\omega_{+}}d\omega_k (p^2-2|p_0|\omega_k)^4 n_B(\omega_k) \quad (4m^2<p^2)\\ \\
& 0\quad (0<p^2<4m^2)\\ \\
& (-i\hbar)\frac{\pi}{4|{\bf p}|}\left(\int_{\omega_{-}}^{\infty} d\omega_k (p^2+|p_0|\omega_{k})^4 n_B(\omega_k)+\int_{\omega_{+}}^{\infty} d\omega_k (p^2-|p_0|\omega_{k})^4 n_B(\omega_k) \right)\quad (p^2<0)\,.
\end{cases}
\end{align}
Finally, we find that the free-thermal part vanishes
\begin{align}
    \mathcal{N}^{0\beta}(p)&=0\,,
\end{align}
and does not contribute to the decoherence.

\subsection{Thermal-thermal part}
In the same way as in the free-thermal part, we can evaluate the thermal-thermal part to obtain
\begin{align}
    \mathcal{N}^{\beta\beta}(p)&= 2{\rm Re}\left[I_{1}^{\beta\beta}(p)-\frac{2}{|\bm{p}|^2}I_{2}^{\beta\beta}(p) + \frac{1}{|\bm{p}|^4}I_{3}^{\beta\beta}(p)\right]\,,
\end{align}
where
\begin{align}
I_{1}^{\beta\beta} =
\begin{cases}
& \frac{2\pi^2}{|{\bf p}|}\int_{\omega_{-}}^{\omega_{+}}d\omega_k n_B(|p_0|-\omega_k)n_B(\omega_k)(\omega_k^2-m^2)^2\quad (4m^2<p^2)\\ \\
& 0\quad (0<p^2<4m^2)\\ \\
& \frac{2\pi^2}{|{\bf p}|}\left(\int_{\omega_{-}}^{\infty} d\omega_k n_B(|p_0|+\omega_k)n_B(\omega_k)(\omega_k^2-m^2)^2 +\int_{\omega_{+}}^{\infty} d\omega_k n_B(|p_0|-\omega_k)n_B(\omega_k)(\omega_k^2-m^2)^2\right)\quad (p^2<0)\,,
\end{cases}
\end{align}
\begin{align}
I_{2}^{\beta\beta} =
\begin{cases}
& \frac{\pi^2}{|{\bf p}|}\int_{\omega_{-}}^{\omega_{+}}d\omega_k (\omega_k^2-m^2)(p^2-2|p_0|\omega_k)^2n_B(|p_0|-\omega_k)n_B(\omega_k) \quad (4m^2<p^2)\\ \\
& 0\quad (0<p^2<4m^2)\\ \\
& \frac{\pi^2}{2|{\bf p}|}\Bigg(\int_{\omega_{-}}^{\infty} d\omega_k (\omega_k^2-m^2)n_B(\omega_k+|p_0|)n_B(\omega_k)(p^2-2|p_0|\omega_k)^2\\
&+\int_{\omega_{+}}^{\infty} d\omega_k (\omega_k^2-m^2)n_B(|p_0|-\omega_k)n_B(\omega_k)(p^2+2|p_0|\omega_k)^2 \Bigg)\quad (p^2<0)\,,
\end{cases}
\end{align}
and
\begin{align}
I_{3}^{\beta\beta} =
\begin{cases}
& \frac{(2\pi)^2}{32|{\bf p}|}\int_{\omega_{-}}^{\omega_{+}}d\omega_k (p^2-2|p_0|\omega_k)^4n_B(|p_0|-\omega_k)n_B(\omega_k) \quad (4m^2<p^2)\\ \\
& 0\quad (0<p^2<4m^2)\\ \\
& \frac{(2\pi)^2}{32|{\bf p}|}\Bigg(\int_{\omega_{-}}^{\infty} d\omega_k n_B(\omega_k+|p_0|)n_B(\omega_k)(p^2-2|p_0|\omega_k)^4\\
&+\int_{\omega_{+}}^{\infty} d\omega_k n_B(|p_0|-\omega_k)n_B(\omega_k)(p^2+2|p_0|\omega_k)^4 \Bigg)\quad (p^2<0)\,.
\end{cases}
\end{align}

Finally, we obtain the thermal-thermal contribution,
\begin{align}
    \mathcal{N}^{\beta\beta}(p)=
    \begin{cases}
        &-\frac{\pi^2}{60p^4}e^{-\beta |p_0|} \sqrt{1-\frac{4m^2}{p^2}}\Big( 17p^8 - 366p^6p_0^2 + 366p^4p_0^4 -16m^4(23p^4+56p^2p_0^2-96p_0^4)\\
    &\quad+8m^2(18p^6+181p^4p_0^2-216p^2p_0^4)\Big) \quad(4m^2<p^2)\,,\\ \\
    &0\quad(0<p^2<4m^2)\,,\\ \\
    & -\frac{\pi^2}{ \beta^5}\frac{p^2}{|{\bm p}|^5}e^{-\sqrt{\left( 1-\frac{4m^2}{p^2}\right)|{\bm p}|^2}\beta}\Big( 4m^2p_0^2\beta^2(1+2p_0^2\beta^2)+\beta^2p^4(1+2p_0^2\beta^2) \\ 
    &\quad-p^2 \Big( 3+7p_0^2 \beta^2 + 4 p_0^4 \beta^4 + 4m^2\beta^2 (1+2p_0^2\beta^2) +3\sqrt{\left( 1-\frac{4m^2}{p^2}\right)|{\bm p}|^2}\beta (1+2p_0^2\beta^2) \Big) \Big)
    \quad(p^2<0)\,,
    \end{cases}
\end{align} 
at low temperature limit.
At high temperature limit, we find
\begin{align}
    \mathcal{N}^{\beta\beta}(p)=
    \begin{cases}
    & -\frac{\pi^2}{6p^2(p^2-p_0^2)^2\beta^2}\sqrt{1-\frac{4m^2}{p^2}}\Big( 11p^8 -46p^6p_0^2 + 82p^4 p_0^4 -44p^2 p_0^6 +4m^2(-5p^6+p^4p_0^2+2p^2p_0^4 + 2p_0^6 )\Big)\\
    & + \frac{\pi^2}{2p_0(-p^2+p_0^2)^{5/2}\beta^2}{\rm Arctanh}\left[ \frac{|{\bm p}|}{|p_0|}\sqrt{1-\frac{4m^2}{p^2}}\right] \Big( p^8 + 4p^6p_0^2 - 20p^4p_0^4+32p^2p_0^6 -16p_0^8\\
    &+16m^4(p^2-p_0^2)^2 -16m^2(p^2-p_0^2)^2(p^4-p^2p_0^2 + p_0^4)
    \Big) \quad(4m^2<p^2)\,,\\ \\
    &0\quad(0<p^2<4m^2)\,,\\ \\
    &\frac{4\pi^2p^4}{3 (-p^2 +p_0^2)^{5/2}\beta^5} -\pi^2\frac{192m^2p^4 -48p^6 -192m^2p^2p_0^2 -384p^4p_0^3}{24(-p^2p_0^2)^{5/2}\beta^3}\\
    &\quad -\frac{\pi^2}{\beta^2} \frac{-96m^2p^4 +24p^6+96m^2p^2p_0^2 + 276 p^4 p_0^2 + 4p^4 \sqrt{1-\frac{4m^2}{p^2}}(-p^2+p_0^2)}{24 (-p^2+p_0^2)^2}\\
    &\quad +\frac{\pi^2}{\beta^2}\frac{96m^4p^4 - 48m^2p^6 + 6p^8 - 192m^4 p^2 p_{0}^2 + 48m^2 p^4 p_{0}^2 + 96m^4 p_{0}^4}{24 |p_{0}| (-p^2+p_{0}^2)^{5/2}}\log\left[ \frac{-p_0 + \sqrt{1-\frac{4m^2}{p^2}}(-p^2+ p_0^2)^{1/2}}{p_0 + \sqrt{1-\frac{4m^2}{p^2}}(-p^2+ p_0^2)^{1/2}}\right]\\
    &\quad +\frac{\pi^2}{\beta^2}\frac{48m^4 p^4 -24 m^2 p^6 + 3p^8 -96m^4p^2p_0^2 -168m^2 p^4 p_0^{2}+48p^6p_0^2 + 48m^4 p_0^4 +192m^2p^2p_0^4 +192p^4 p_0^4}{24|p_0|(-p^2+p_0^2)^{5/2}}\log\left[\frac{1+\beta|p_0|}{1-\beta|p_0|} \right]\\
    &\quad +\frac{\pi^2}{\beta^2}\frac{-192m^2p^4p_{0}^2+48p^6 p_{0}^2 + 192m^2 p^2 p_{0}^4 +192p^4 p_{0}^4}{24p_0 (-p^2 +p_{0}^2)^{5/2}}\log\left[\beta^2\left\{-p^2 -\frac{4m^2}{p^2}(-p^2+p_{0}^2) \right\}\right]\\
    &\simeq \frac{4\pi^2p^4}{3 (-p^2 +p_0^2)^{5/2}\beta^5}   \quad(p^2<0)\,.
    \end{cases}
\end{align}

\bibliography{bib}

\end{document}